%% file: Edges_v2.tex
\documentclass[useAMS,usenatbib]{mn2e}
\usepackage{amssymb,amsmath,epsfig,times, natbib,color,graphicx}
\usepackage{hyperref}
\pdfoutput=1
\hypersetup{
    colorlinks=false,
    linkcolor=blue,
    filecolor=blue,      
    urlcolor=blue,
}

\title[Edges with Chandra and XMM]{Applications for edge detection techniques using Chandra and XMM-Newton data: galaxy clusters and beyond}
\author[S. A. Walker et al.]{S. A. Walker,$^1$\thanks{Email: 
    swalker@ast.cam.ac.uk} J. S. Sanders$^2$ and A. C. Fabian$^1$ \\
  $^1$Institute of Astronomy, Madingley Road, Cambridge CB3 0HA \\
  $^2$Max-Planck-Institute fur extraterrestrische Physik, 85748 Garching, 
Germany  \\
  \\
    \\
   \\
   \\
}
\date{}

\begin{document}

\maketitle

\begin{abstract}
The unrivalled spatial resolution of the Chandra X-ray observatory has allowed many breakthroughs to be made in high energy astrophysics. Here we explore applications of Gaussian Gradient Magnitude (GGM) filtering to X-ray data, which dramatically improves the clarity of surface brightness edges in X-ray observations, and maps gradients in X-ray surface brightness over a range of spatial scales. In galaxy clusters, we find that this method is able to reveal remarkable substructure behind the cold fronts in Abell 2142 and Abell 496, possibly the result of Kelvin Helmholtz instabilities. In Abell 2319 and Abell 3667, we demonstrate that the GGM filter can provide a straightforward way of mapping variations in the widths and jump ratios along the lengths of cold fronts. We present results from our ongoing programme of analysing the Chandra and XMM-Newton archives with the GGM filter. In the Perseus cluster we identify a previously unseen edge around 850 kpc from the core to the east, lying outside a known large scale cold front, which is possibly a bow shock. In MKW 3s we find an unusual `V' shape surface brightness enhancement starting at the cluster core, which may be linked to the AGN jet. In the Crab nebula a new, moving feature in the outer part of the torus is identified which moves across the plane of the sky at a speed of $\sim$0.1$c$, and lies much further from the central pulsar than the previous motions seen by Chandra. 
 
\end{abstract}


\section{Introduction}
Thanks to its unsurpassed sub-arcsecond spatial resolution, the X-ray images obtained with the Chandra X-ray observatory have transformed our understanding of many astrophysical processes. Of particular interest are sharp surface brightness discontinuities. In galaxy clusters, phenomena such as cold fronts due to gas sloshing and merging (e.g. \citealt{Markevitch2000}, \citealt{Tittley2005}, \citealt{Markevitch2007}, \citealt{Owers2009}, \citealt{Ghizzardi2010}), shocks due to merging activity (e.g. \citealt{Markevitch2002}, \citealt{Russell2012}), and cavities in the intracluster medium (ICM) resulting from powerful AGN feedback (\citealt{Boehringer1993}, \citealt{McNamara2000}, \citealt{Fabian2006}) leave a myriad of such surface brightness features in X-ray images. Galaxy clusters also feature X-ray surface brightness features which are much more subtle, such as sound waves (\citealt{Fabian2006}) and substructure behind cold fronts (\citealt{Werner2016}, \citealt{Sanders2016}), with variations of only a few percent. Our ability to unravel the physics of the intracluster medium is limited by our ability to resolve features which have low contrast in X-ray images.     

Recently, in \citet{Sanders2016} and \citet{Sanders2016GGM}, we have applied a Gaussian gradient magnitude (GGM) filter to map the complex X-ray surface brightness features in Chandra images of nearby galaxy clusters which have deep observations, namely the Perseus cluster, M87 and the Centaurus cluster. This technique has dramatically improved the clarity with which we can identify structure in X-ray images, and promises to significantly increase the scientific return from Chandra observations of a wide range of phenomena. 

Here we explore applications of this GGM filter method to the archive of Chandra and XMM-Newton observations. We investigate how it can help improve our understanding of the physics occurring in a wide range of objects, and how it can be used as a discovery tool for identifying previously unseen features in existing data. One immediate application is in the study of cold fronts in galaxy clusters. These are sharp surface brightness edges seen in clusters, with cooler gas on the brighter, denser side and hotter gas on the lower density side (the opposite situation to a shock). Two main types of cold fronts have been observed in clusters; those in clusters which retain a relaxed morphology (e.g. RXJ 1720+26, \citealt{Mazzotta2001}; A1795, \citealt{Markevitch2001}; A2029, \citealt{Clarke2004}), and those in clusters which have clearly undergone significant merging activity (e.g. A3667, \citealt{Vikhlinin2001}; the Bullet cluster, \citealt{Markevitch2002} and NGC 1404 in the Fornax cluster, \citealt{Machacek2005}). In the first case for relaxed clusters, the cold fronts are believed to be produced due to sloshing motions of cluster cool cores, which are induced by off axis minor mergers (\citealt{Ascasibar2006}), leading to a spiral pattern of alternating cold fronts at increasing radius. In the second case, the cold fronts are produced by the merging motion of the cool core of through the surrounding hotter ICM.

The sharpness of these edges (they are much narrower than the Coulomb mean free path) is possibly the result of a considerable magnetic field draped across their surfaces (e.g. \citealt{Lyutikov2006}, \citealt{Asai2007}, \citealt{Dursi2008}). This would severely inhibit transport processes and prevent cold fronts from disintegrating through Kelvin-Helmholtz instabilities (KHIs), which would otherwise be expected to develop on relatively short timescales (\citealt{Chandrasekhar1961}). Numerical simulations (e.g. \citealt{ZuHone2011}, \citealt{Roediger2013}),  have provided predictions for the morphology of KHIs that could be expected behind cold fronts, and how these vary as a function of the viscosity of the ICM (see \citealt{ZuHone2016} for a review). Observational identification of instabilities in cold fronts is challenging with Chandra owing to the low contrast of these features against the bright X-ray emission from the ICM, and provides an excellent application for the GGM filter. \citet{Roediger2013} have previously applied GGM filtering to examine simulations of gas sloshing and cold fronts in galaxy clusters. Mapping the way the widths of cold fronts vary along their length is also challenging, and in this work we explore how well GGM filtered images can provide simple ways of mapping variations of cold front width and density jump ratio.

GGM filtered images also provide a powerful tool for discovering new features in X-ray images. Their power stems from the fact that the gradient can be mapped on multiple spatial scales scales, which can then be combined to produce one single image. This is a significant advance over previous techniques such as unsharp masking, in which the filtering has to be manually tuned depending on the size of the features being studied. GGM filtering is therefore well suited to the analysis of large sets of data, such as the Chandra and XMM-Newton archives, as it requires little initial configuration. Here we explore and quantify the improvement in contrast of a myriad of surface brightness features in galaxy clusters achieved with the filter. We then present some new discoveries from our ongoing programme to analyse the entire Chandra and XMM-Newton archive.

Our results are structured as follows. In section \ref{substructuremapping} we demonstrate the GGM filter's ability to reveal possible substructure behind the cold fronts in Abell 2142 and Abell 496. In section \ref{widthmapping} we show that, in certain circumstances, the filter can be used to produce simple maps of the variation of the width or density jump ratio along the length of cold fronts, allowing the differences between cold fronts to be better understood. Section \ref{soundwaves} explores how well the filter can reveal ripples in galaxy clusters using simulated images of an idealised cluster with similar ripple features to those observed in the Perseus cluster. The amount by which the filter improves the contrast of features in galaxy clusters is quantified in section \ref{quantifyingimprovement}, in which we compare the contrast of a wide variety of features in Chandra images of clusters before and after filtering. Finally, in section \ref{discoverytool} we present three exciting new features we have identified in our ongoing programme to analyse the entire Chandra and XMM-Newton archive with the GGM filter.

We use a standard $\Lambda$CDM cosmology with $H_{0}=70$  km s$^{-1}$
Mpc$^{-1}$, $\Omega_{M}=0.3$, $\Omega_{\Lambda}$=0.7. All errors unless
otherwise stated are at the 1 $\sigma$ level. The figures in this paper are best viewed on a computer screen.

\section{Data}

The data for the various sources we study are tabulated in table \ref{alldata} in Appendix A. We obtained exposure corrected, background and point source subtracted Chandra images in the 0.7-7.0 keV band using the methods described in \citet{Walker2014_A1795}. In short, we reprocessed each events file using \textsc{chandra\_repro} in CIAO 4.8. The routine \textsc{lc\_sigma\_clip} was used to identify periods where the count rate differed from the mean by more than 2$\sigma$, allowing periods of flaring to be removed. \textsc{reproject\_obs} was used to reproject the events files, and images and exposure maps in the 0.7-7.0 keV band were extracted using \textsc{flux\_obs}. When a particular object had multiple exposures these were combined together, weighting by their exposure maps to produce stacked images. To remove the background we used the CIAO script \textsc{acis\_bkgrnd\_lookup} to identify suitable blank sky background fields for each events file, which were reprojected to match the coordinate system of the observations. We scaled the blank sky backgrounds so that their hard band (10-12keV) count rates (where the signal is just due to the particle background as the effective area is zero) matched those of the observations before subtracting. Point sources were identified using \textsc{wavdetect} (and verified by eye) and removed from the images. The pixels values of the point sources were replaced with values obtained by randomly sampling the pixels in the immediate vicinity of the point sources, thus allowing a continuous image to be obtained. 

In section \ref{Perseusedge} we examine an XMM-Newton mosaic of the Perseus cluster. To produce this XMM-Newton mosaic we used the XMM Extended source Analysis (ESAS) software\footnote{ftp://legacy.gsfc.nasa.gov/xmm/software/xmm-esas/xmm-esas.pdf} (\citealt{Snowden2008}), following the methods described in depth in \citet{Walker2013_CentaurusXMM}.

\section{Gaussian Gradient Magnitude filter}

We have introduced the use of the Gaussian Gradient Magnitude filter on Chandra data in \citet{Sanders2016} and \citet{Sanders2016GGM}, in which we describe the filtering process in detail. We use the implementation of GGM available in \textsc{SCIPY},  (\citealt{SciPy}, http://scipy.org/). This filter calculates the gradient in an image assuming Gaussian derivatives. The image is convolved with the gradient of a one dimensional Gaussian, and the gradient in the image is calculated in two directions, which are then combined to produce a final image of the gradient distribution. Different sized Gaussians are used to  map the gradients on different spatial scales, with a smaller width needed to map small scale gradient variations, as demonstrated in \citet{Sanders2016GGM}. The GGM images we present are produced by combining together images filtered on the lengths scales of $\sigma=$1, 2, 4, 8 and 16 pixels, weighting each of these using the radial weighting scheme discussed in \citet{Sanders2016GGM}. The method allows equal emphasis to be placed on sharp features in cluster cores, and broader features at larger cluster radii. 

\section{Mapping Structure Behind Cold fronts in galaxy clusters}
\label{substructuremapping}
\subsection{Abell 2142}
\begin{figure*}
  \begin{center}
    \leavevmode
    \hbox{\includegraphics[width=0.6\linewidth]{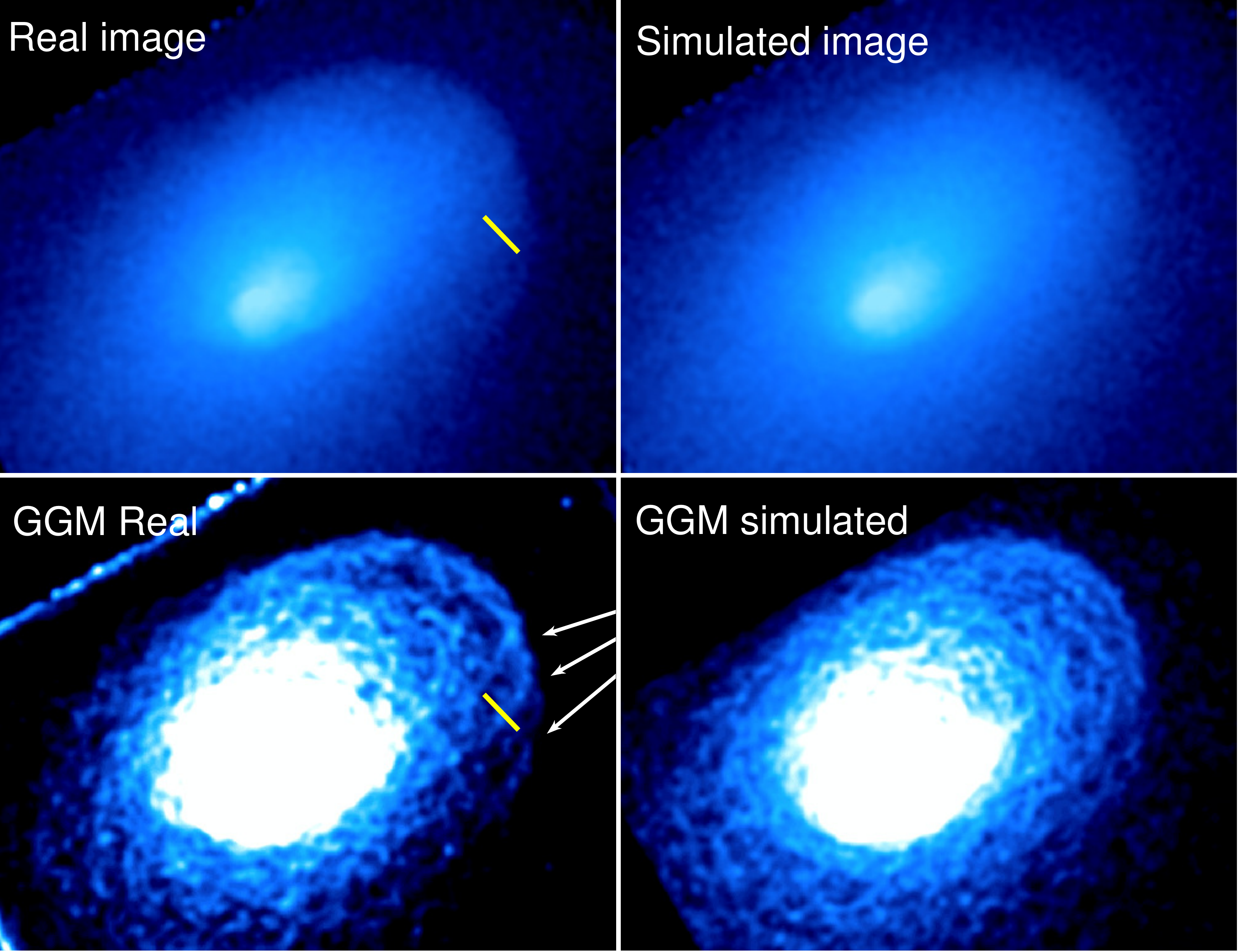}
     \includegraphics[width=0.4\linewidth]{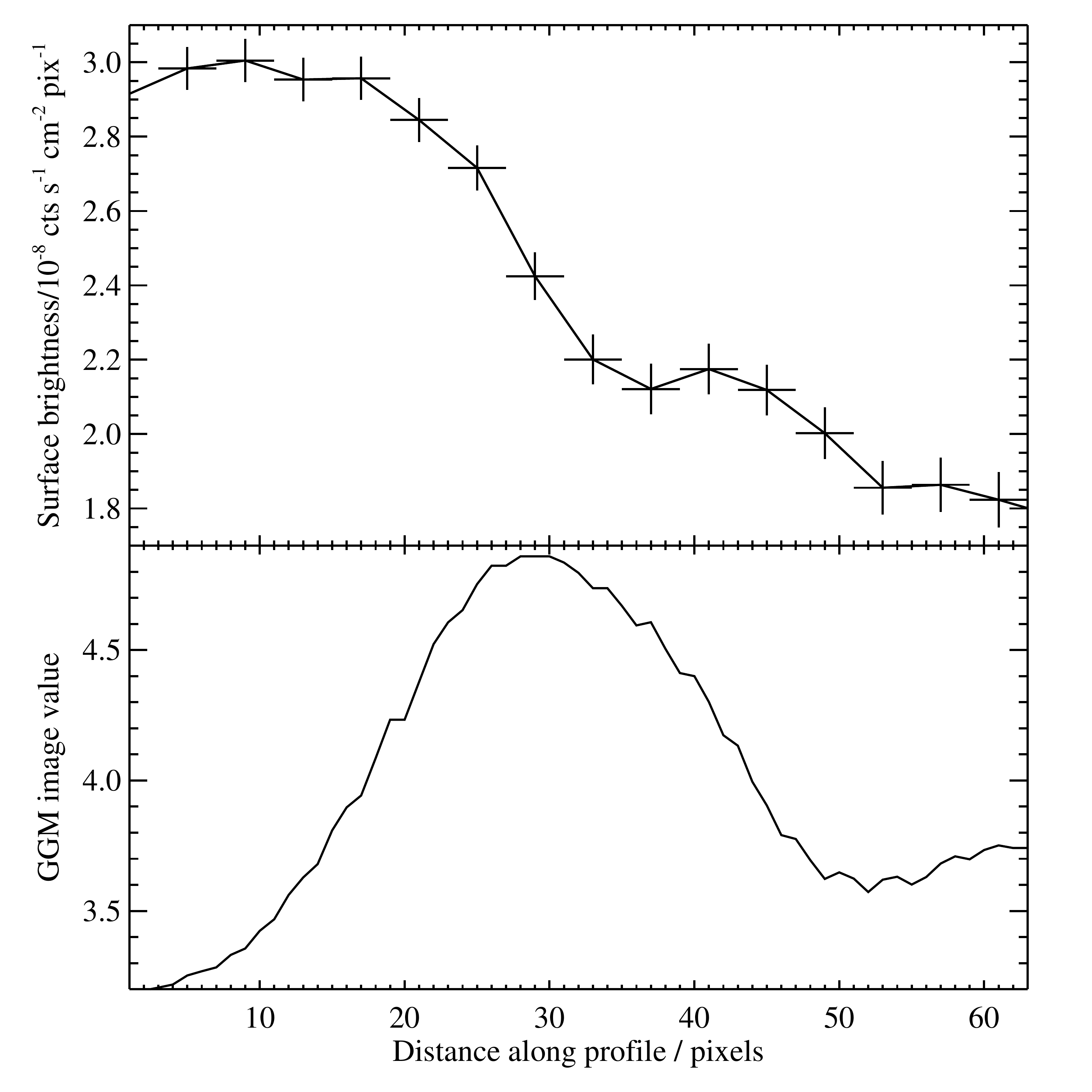}
               
        }

      \caption{Top left: 0.7-7.0 keV band Chandra image of A2142. Bottom left: GGM filtered version of the top right image, showing linear features behind the cold front marked with white arrows. Top centre: simulated Chandra image of Abell 2142 obtained by fitting the real image with 30 ellipsoids to reconstruct its large scale shape and surface brightness edges. The effect of running the GGM filter on this simulated image is shown in the bottom centre panel, in which we see no linear features behind the cold front, showing that they are not artefacts of the GGM filtering process. Right: Plots comparing profiles along the yellow lines shown in the left hand panels from the original image (top) and the filtered image (bottom). These images are best viewed on a computer screen. }
      \label{A2142_images}
  \end{center}
\end{figure*}

One of Chandra's first major discoveries in galaxy clusters was the prominent cold front in Abell 2142 (\citealt{Markevitch2000}, \citealt{Owers2011}, \citealt{Rossetti2013}), shown in the top left panel of Fig. \ref{A2142_images}, which is now believed to have formed due to the sloshing of the cluster core in response to an off axis merger. Applying the GGM filter to the 0.7-7.0keV band image of A2142 reveals linear substructure extending from the edge of the main cold front, highlighted by the white arrows in the bottom left panel of Fig. \ref{A2142_images}. These features bear a striking resemblance to the features found behind the cold front in the Centaurus cluster using the same filtering technique in \citet{Sanders2016} (features L1, L2 and L3 in their figure 7). Similar linear features have also been found behind the cold front in the Virgo cluster by \citet{Werner2016}.

To test that these features a real, and not just a relic of the GGM filtering, we performed the following test. We simulated images of Abell 2142 by fitting the real image with 30 ellipsoids (which reproduces the large scale sloshing features and edges), and then used the exposure maps to generate 100 Poisson realisations, one of which is shown in the top centre panel of Fig. \ref{A2142_images}. We then applied the GGM filter to these simulated images, one of which is shown in the bottom centre panel of Fig. \ref{A2142_images}. We see that the linear features observed in the filtered real data are not present in the filtered simulated data. In all of the 100 Poisson realisations we simulated, no linear features were present. This confirms that these features are genuinely present in the real data, and are not just an artefact of the GGM filtering process.

For further verification, in the plots in the right hand panel of Fig. \ref{A2142_images} we compare profiles across the linear features in the GGM filtered images with profiles across the same regions in the original images. These profiles are taken along the yellow lines shown in the left panels of Fig. \ref{A2142_images}. We see that the surface brightness in the real image fluctuates, with a region of steeper decline between two regions of shallow decline. The profile across the GGM image shows this as a prominent peak where the surface brightness decline is greatest. The variation in surface brightness is rather subtle, at the level of $\approx$8 percent, and can be much seen much more clearly in the filtered image, where the contrast is around five times greater at $\approx$45 percent. 

These features are similar to what has been seen in simulations of Kelvin Helmholtz Instabilities (KHIs) at cold fronts in \citet{Roediger2013} (see their figure 8). In principle, if these structures are KHIs, their size and morphology can be compared to simulations with varying levels of viscosity to put constraints on the viscosity of the ICM at cold fronts.

Another possibility, described in \citet{Werner2016}, is that these linear features are the result of the amplification of the cluster magnetic field between them due to gas sloshing. In this process, the amplified magnetic field provides a considerable magnetic pressure ($\sim$5-10 per cent of the thermal pressure) which decreases the gas density  in linear regions behind the cold front, causing them to be fainter in X-rays. The linear features seen in X-rays therefore become visible because the X-ray surface brightness of the regions between them is suppressed.

\subsection{Abell 496}
\begin{figure*}
  \begin{center}
    \leavevmode
    \hbox{
    \includegraphics[width=0.6\linewidth]{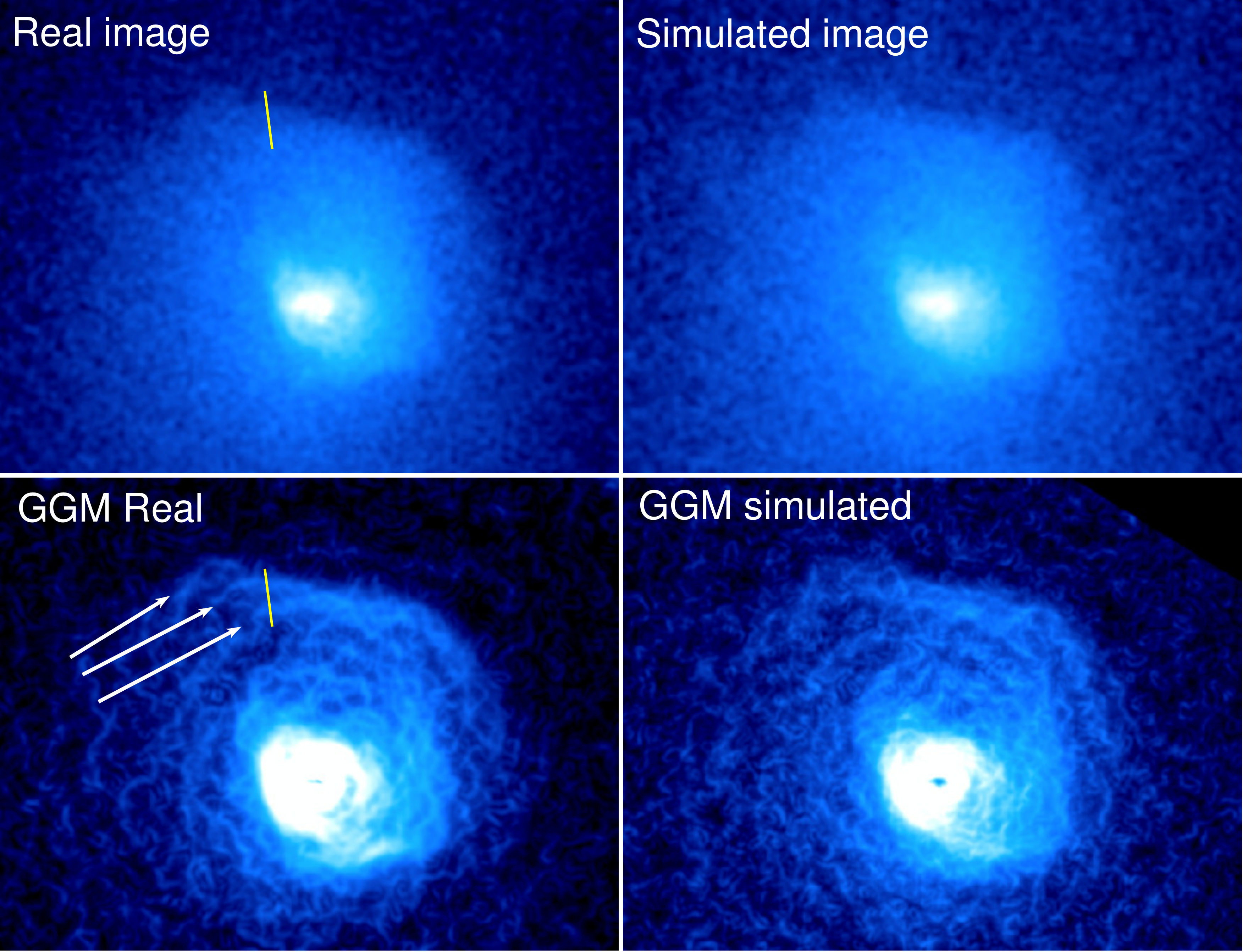}
     \includegraphics[width=0.4\linewidth]{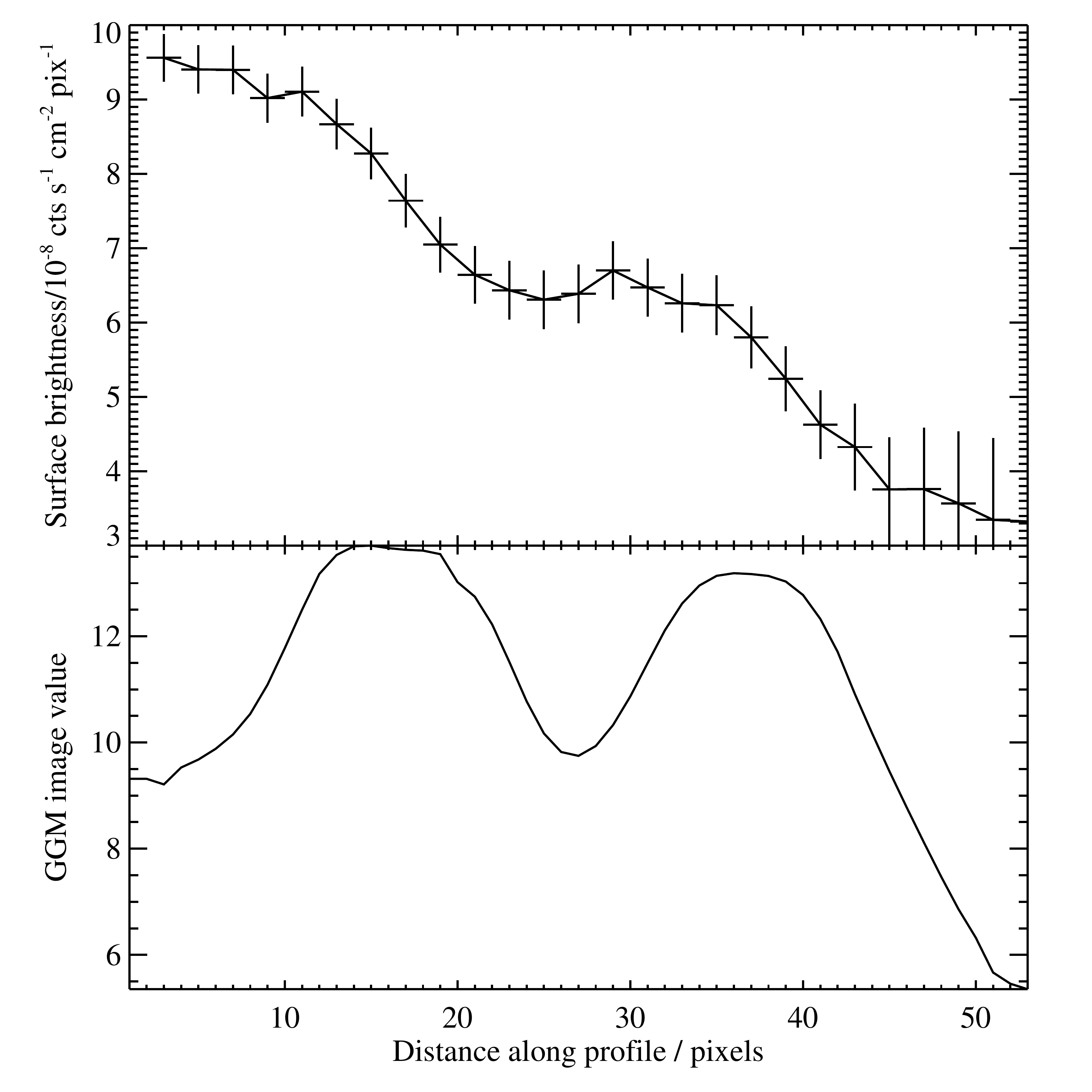}
       
        }

      \caption{Same as Fig. \ref{A2142_images}, but this time for Abell 496. Linear features in the GGM filtered map of the real data are marked by white arrows in the bottom left hand panel, and are not present in the filtered simulated image (bottom centre panel), which again confirms that these are real features and not relics of the GGM filtering process. The right hand plots compare profiles taken along the yellow lines shown in the left hand panels for the original and filtered images. These images are best viewed on a computer screen.}
      \label{A496_images}
  \end{center}
\end{figure*}

Abell 496 is another galaxy cluster with a prominent cold front (\citealt{Dupke2003}), shown in the Chandra images in Fig. \ref{A496_images}. \citet{Roediger2012} identified possible substructure behind this cold front using various X-ray surface brightness profiles along strips across the cold front. Our GGM images allow the substructure to be mapped with great visible clarity. As shown in the GGM image in the bottom left hand panel of Fig. \ref{A496_images}, we see linear features behind the cold front (marked by the white arrows), similar to those we found in A2142. As with A2142, we repeat the test of potential artefacts from the GGM filter by running it on simulated data obtained by fitting the real data with 30 ellipsoids to model its large scale surface brightness distribution. These are shown in the top and bottom centre panels of Fig. \ref{A496_images}, in which we  do not see the linear features, again confirming that they are not relics of the GGM filtering.

Again, in the right hand panel of Fig. \ref{A496_images} we compare a profile across the linear features seen in the GGM image with a profile across the same region (shown by the yellow line in the left hand panels) in the original Chandra image. The peaks in the GGM image correspond to regions of steeper surface brightness decline. The surface brightness features are again very subtle, at the level of $\approx$ 5-7 percent in the original image, and the contrast is much higher in the filtered images, reaching a level of $\approx$40-50 percent. 

\begin{figure*}
  \begin{center}
    \leavevmode
    \vbox{
   \hbox{
       \includegraphics[width=0.7\linewidth]{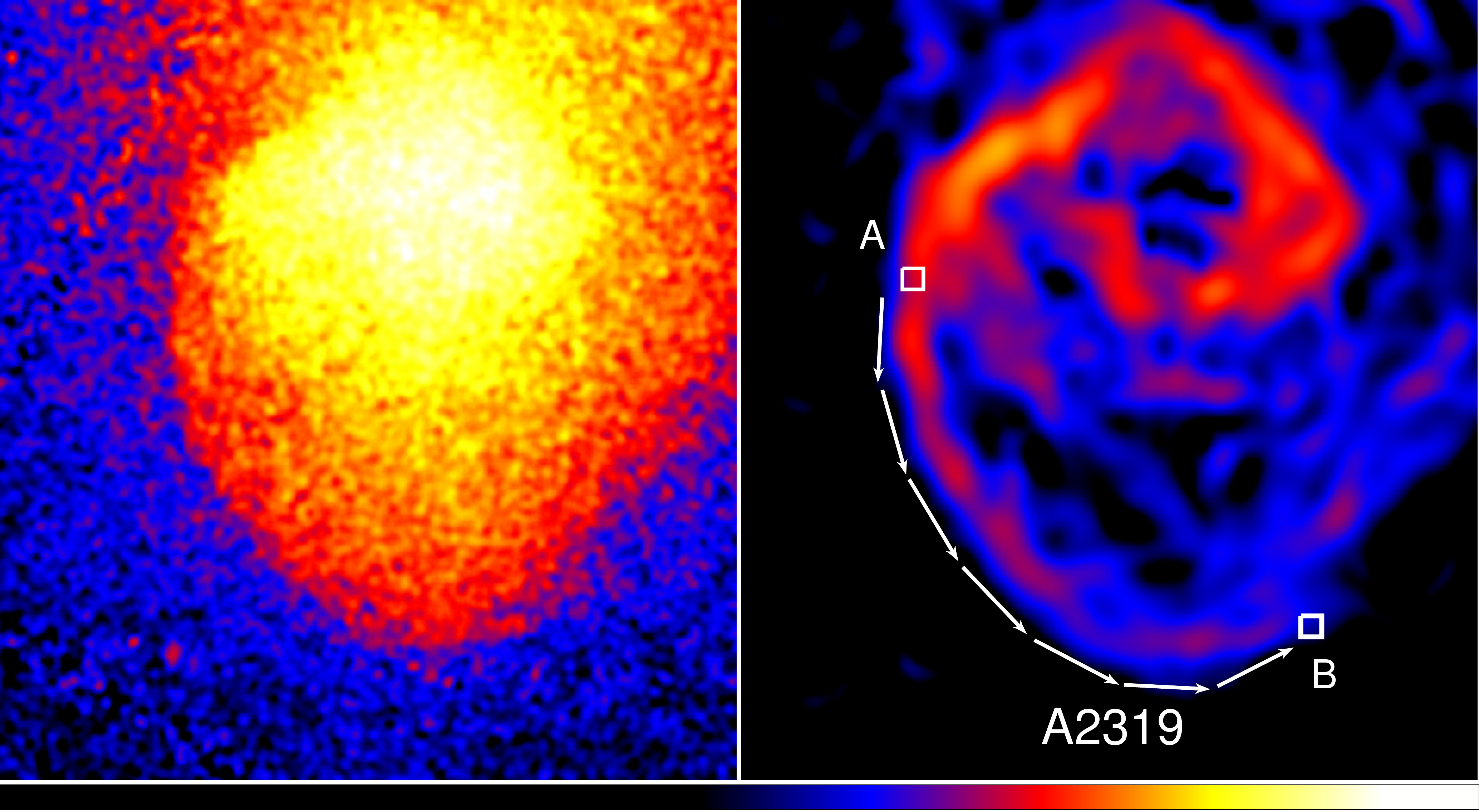}
     \includegraphics[width=0.3\linewidth]{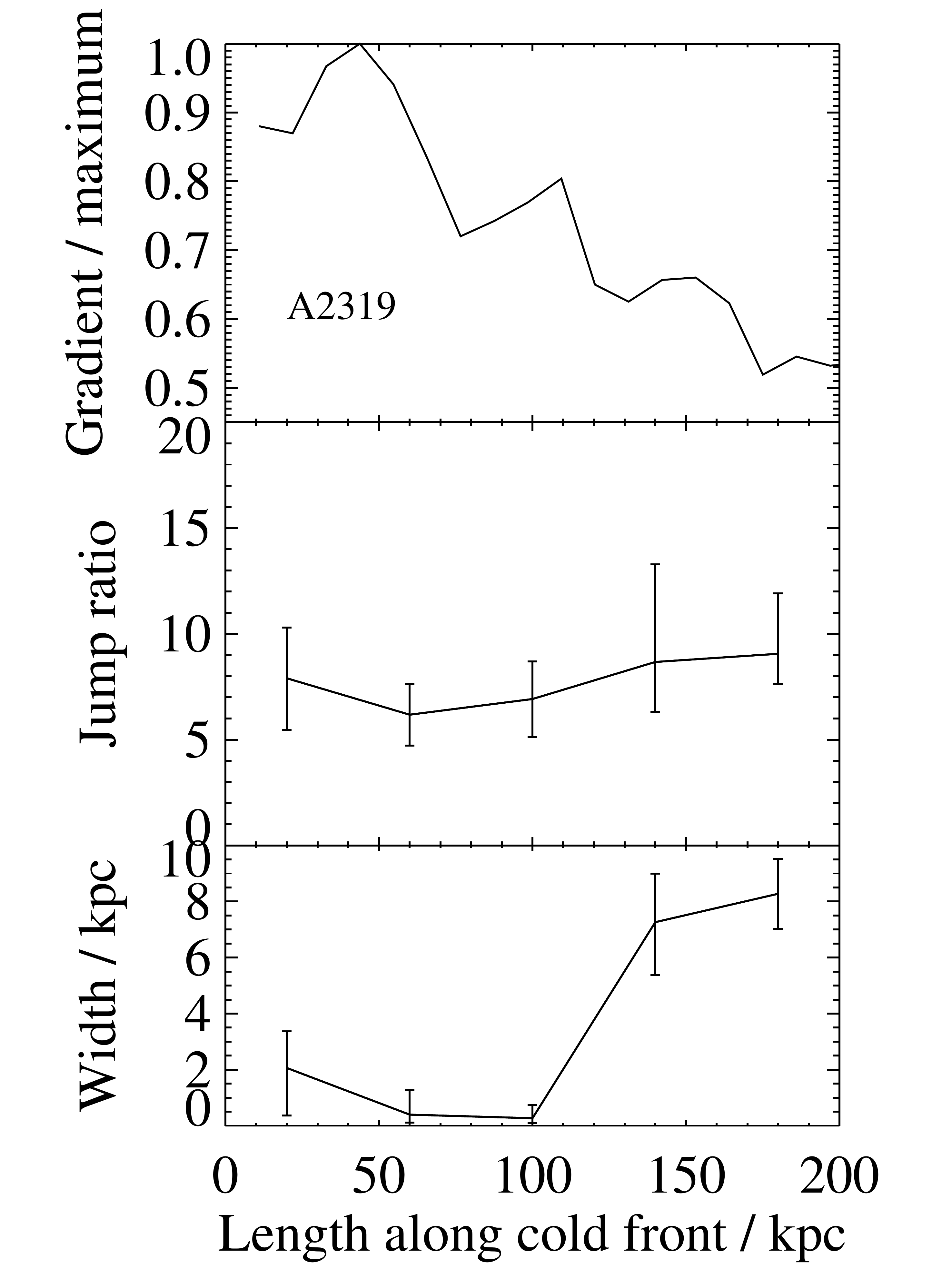}
          }
       \hbox{
          \includegraphics[width=0.7\linewidth]{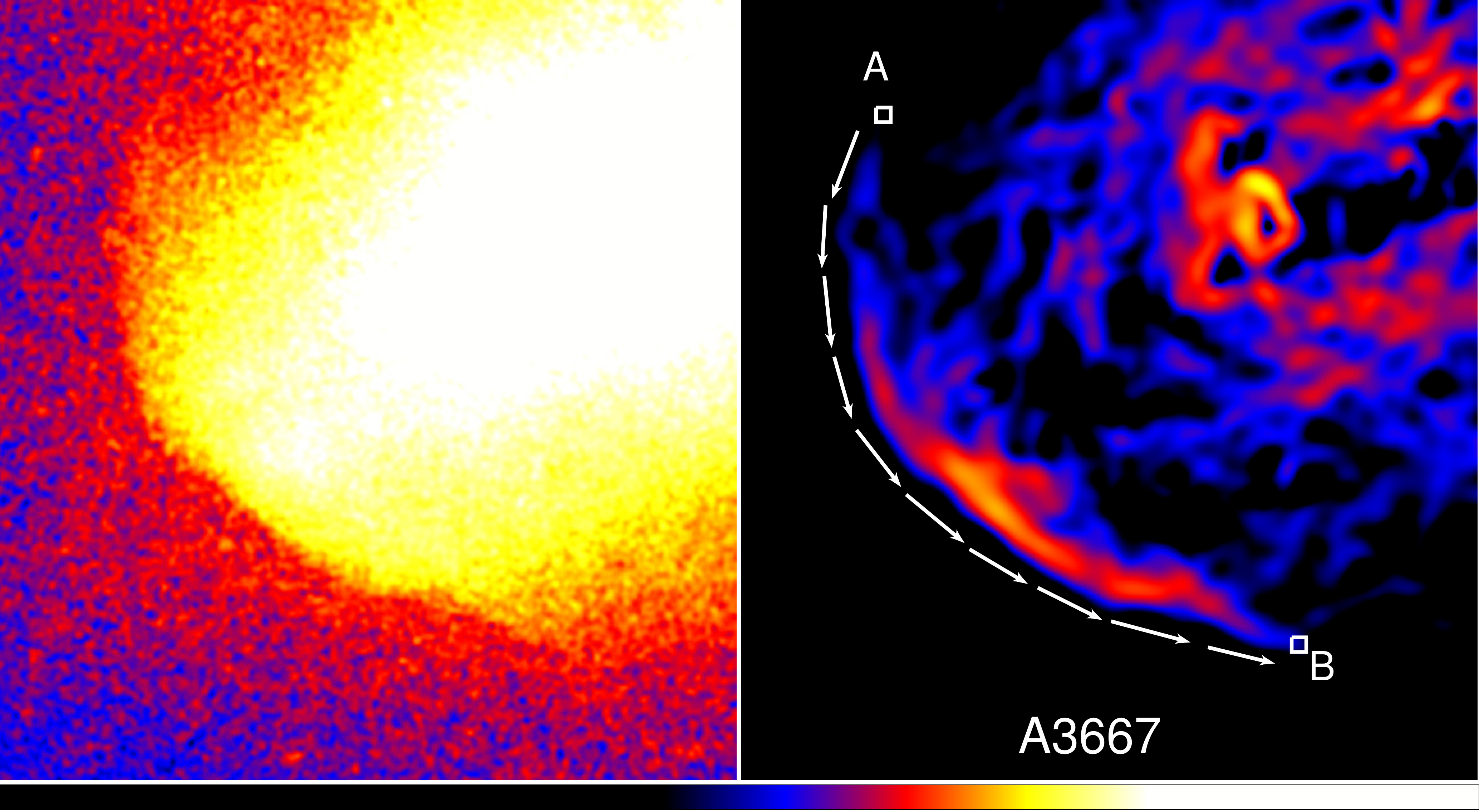}
     \includegraphics[width=0.3\linewidth]{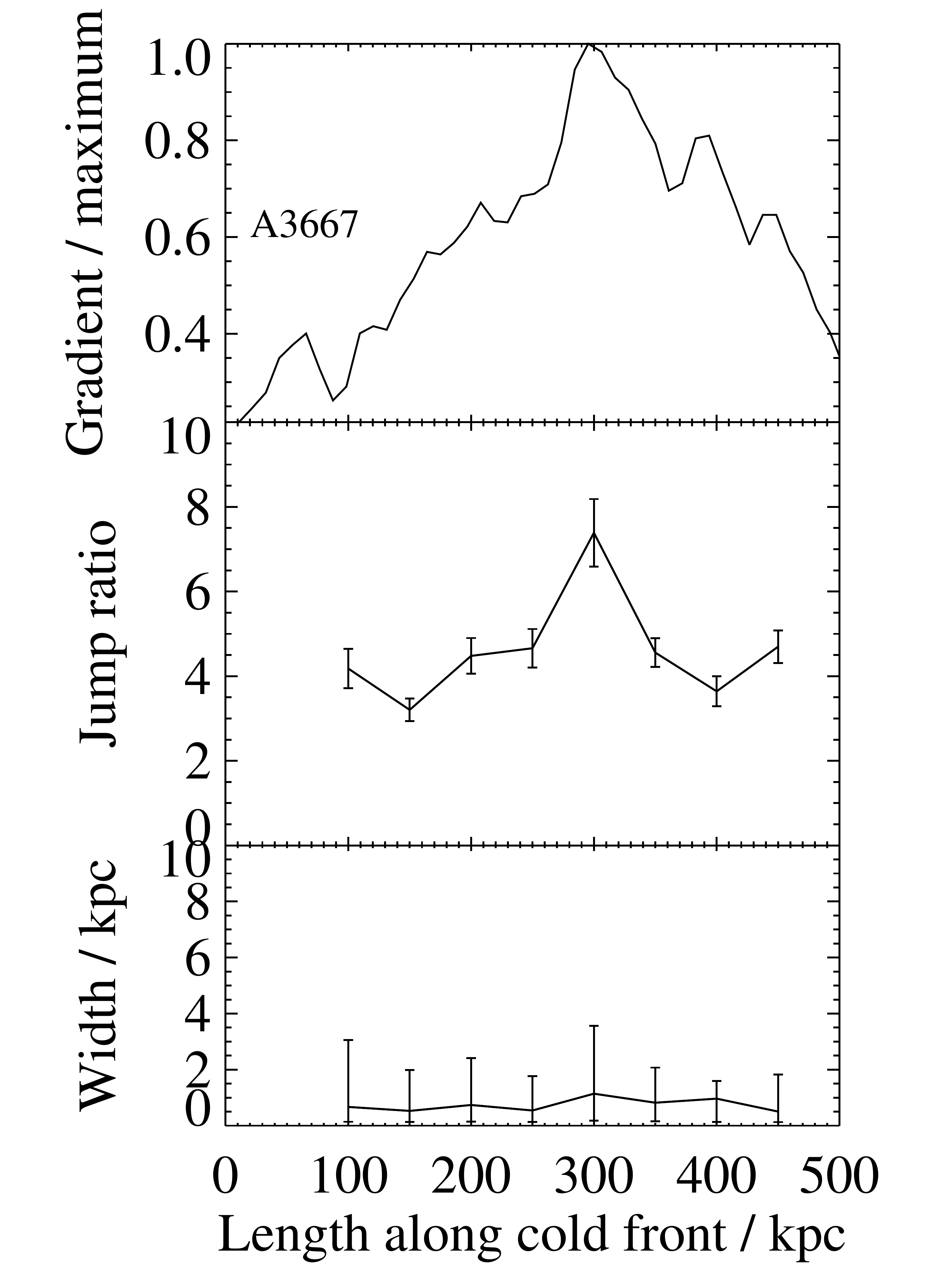}
             
    }}
      \caption{Exploring gradient variations along the cold fronts in A2319 (top row) and A3667 (bottom row). For each cluster the left panel shows the Chandra 0.7-7.0 keV band image, while the centre panel shows the GGM filtered image with a smoothing scale of 8 pixels. The colour bars correspond to the GGM images. The plots in the right most panel show, from top to bottom, the variation in the gradient along the cold fronts obtained from the GGM images, and the variation in the jump ratio and cold front width obtained from surface brightness profile fitting. For these plots we plot the variation of these quantities along the length of the cold fronts, starting from position `A' and following the directions of the arrows to position `B'.    }
      \label{vary_width_realdata}
  \end{center}
\end{figure*}

\begin{figure*}
  \begin{center}
    \leavevmode
   \hbox{
   
             \includegraphics[width=0.7\linewidth]{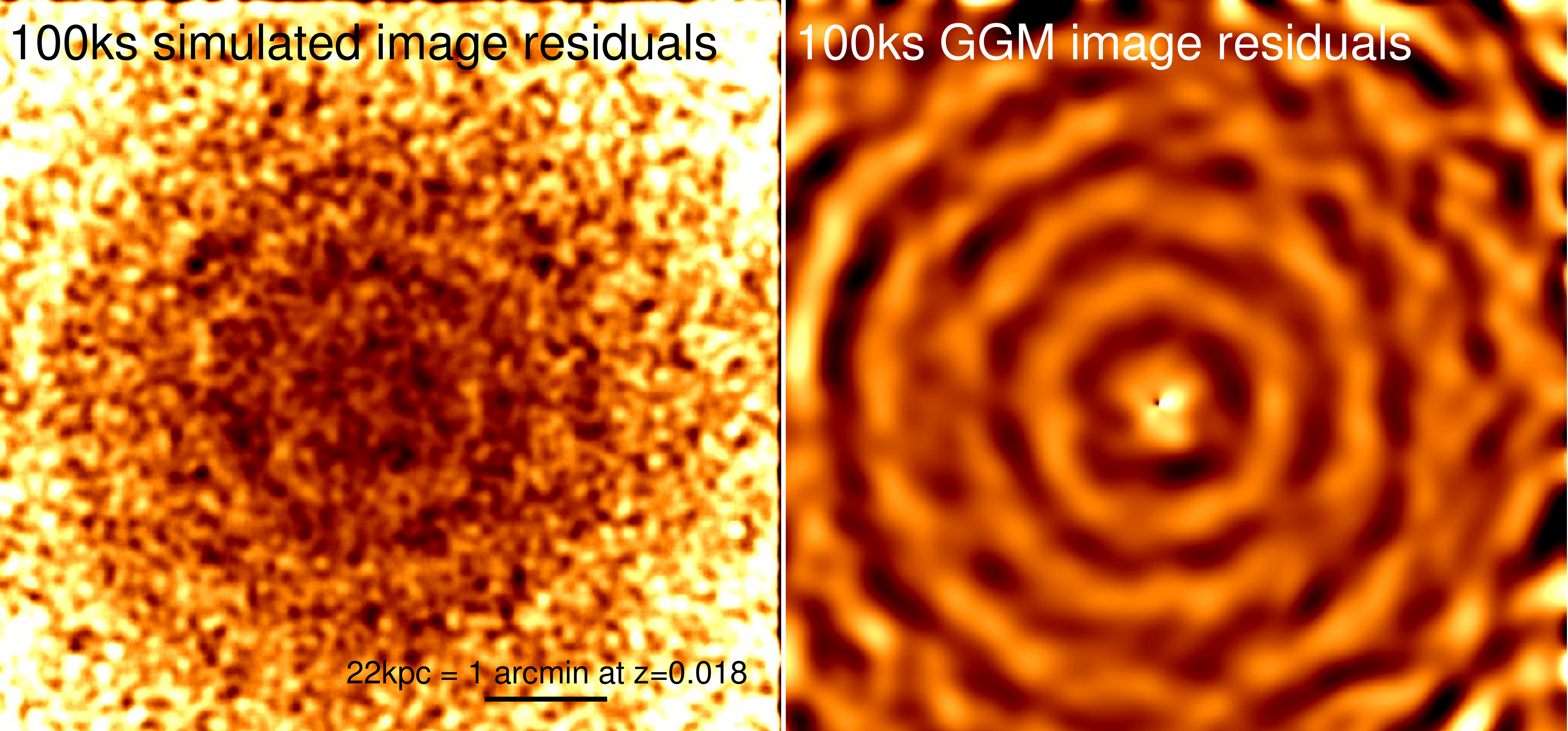}
       
}
        \hbox{
 \includegraphics[width=0.35\linewidth]{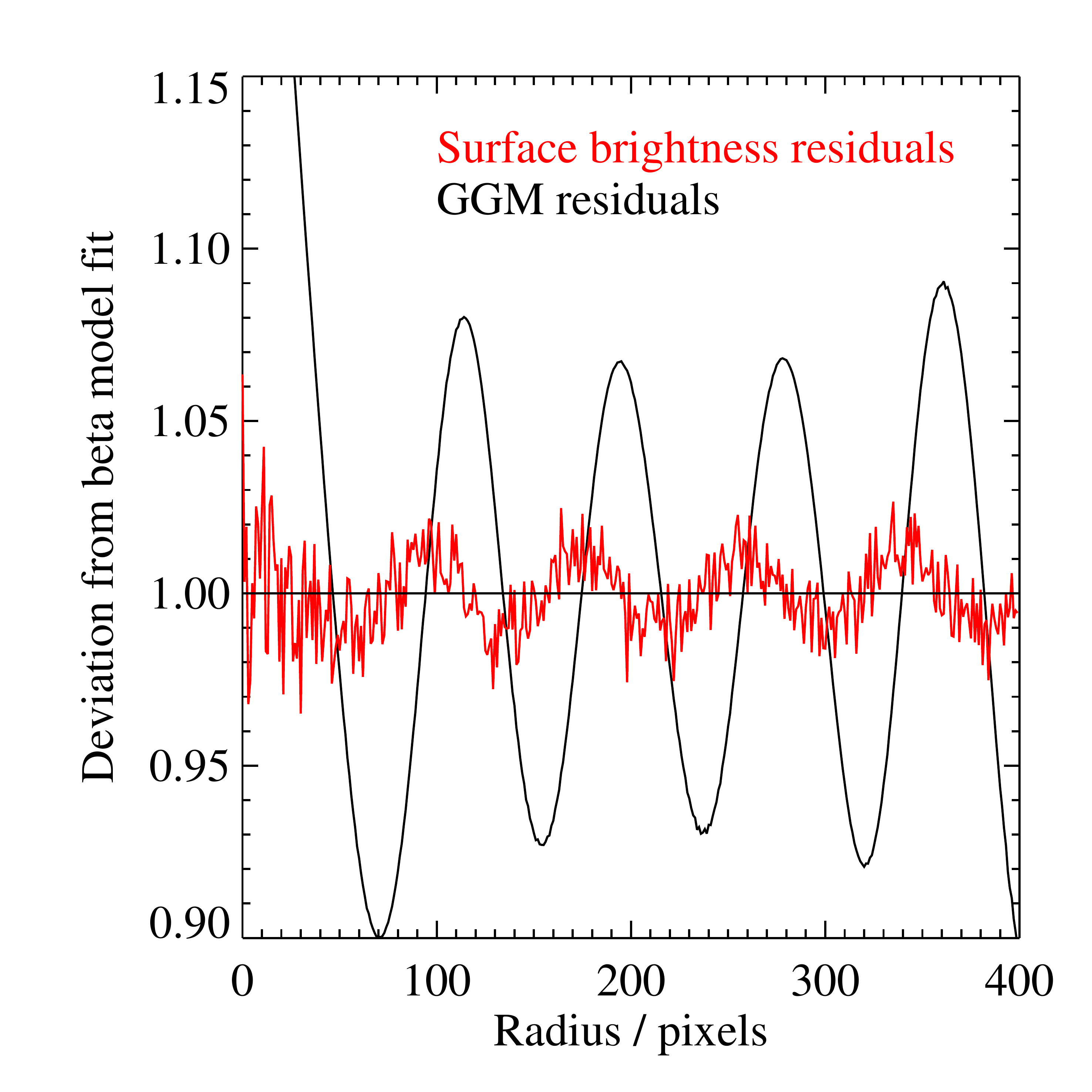}                                     \includegraphics[width=0.4\linewidth]{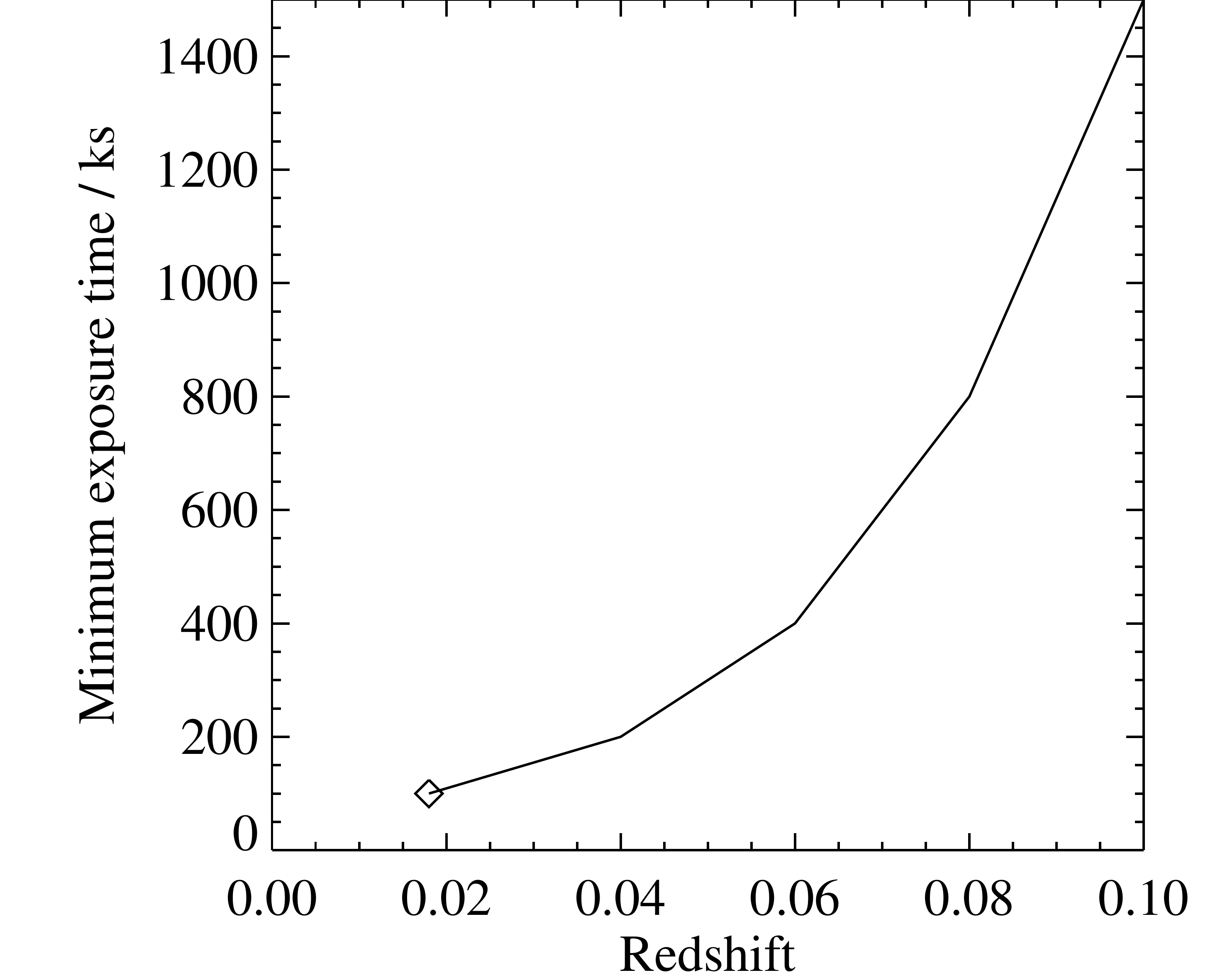}        
                 }      
      \caption{Exploring the detectability of sound waves in galaxy clusters using the GGM filter on simulations. Top left panel: A simulated 100ks observation of a Perseus-like cluster divided by the best fitting beta model, into which we have added ripples with the same magnitude and wavelength as those observed in Perseus.
Top right panel: GGM filter residual image for the same 100ks simulation image, more clearly showing ripple structure. Bottom left: Comparing the profiles of the surface brightness residuals (red) with the GGM residuals (black). Bottom right: Plot of minimum observing time needed to observe ripples of the same characteristics as those in Perseus with increasing redshift.      }
      \label{Soundwavefig}
  \end{center}
\end{figure*}

\section{Mapping the widths of cold fronts}
\label{widthmapping}
\subsection{A2319 and A3667}
One immediate application for the GGM method is to allow us to easily map variations in the widths of cold front in galaxy clusters. Often the variation of this width is very subtle, and not obvious from the raw Chandra images, instead requiring a lengthy process of dividing the edge into numerous sectors, measuring surface brightness profiles across each sector, and fitting the profiles with a model (e.g. \citealt{Werner2016}, \citealt{Sanders2016}). 

To demonstrate this, we examine two prominent, nearby cold fronts in the cluster A2319 (\citealt{OHara2004}, \citealt{Govoni2004}) and A3667 (\citealt{Vikhlinin2001}). The GGM filtered images are shown in Fig. \ref{vary_width_realdata}, from which it is clear that the behaviour of the gradient along the length of the cold fronts is markedly different in the two systems. In A2319 the gradient peaks at the starting point `A', and declines gradually along the cold front as we move to point `B'. This is shown graphically in the top right panel of Fig. \ref{vary_width_realdata}. In A3667 the gradient peaks roughly in the middle of the cold front, and declines on either side, as also shown graphically in the top right hand panel. 

The gradient at the cold front depends on two factors; the width of the cold front and the size of the density drop. A larger density drop leads to a larger gradient, as does a narrower width. To understand the gradient variations seen in the GGM images, we divide the cold fronts into sectors and fit their surface brightness profiles with a model consisting of two power-laws separated by a jump, which is convoluted with a Gaussian to model the width of the surface brightness drop. This same technique is commonly used for measuring the width of cold fronts, and has been used in \citet{Sanders2016} and \citet{Werner2016} to measure variations in width along the  cold fronts in the Centaurus and Virgo clusters respectively. We show the results of these fits in the bottom two panels for the right hand plots for both clusters, which show how the width and the jump ratio vary along the cold fronts. Here the jump ratio is the ratio of the emissivity inside the cold front to that outside. 

For A2319 we see that the jump ratio is remarkably uniform along the length of the cold front, and that the observed decrease in gradient from `A' to `B' is driven by the marked increase in the width. This shows that, for cold fronts with a uniform jump ratio, the GGM images provide a way of mapping variations in cold front width in a way that is far less time consuming than a detailed surface brightness profile fitting method.

In A3667, we see that the width is essentially constant along the whole length of the cold front, and that the gradient variation is driven by changes in the jump ratio, which is highest 300 kpc along the cold front.

\section{Sound waves in galaxy clusters}
\label{soundwaves}
Another application of GGM filtering is the study of sound wave ripples in galaxy clusters. First identified in 200ks observations of the Perseus cluster (\citealt{Fabian2003}), evidence for ripples due to sound waves has since been found in the Centaurus cluster (\citealt{Sanders2008}) and A2052 (\citealt{Blanton2009}). The dissipation of these sound waves provides a mechanism through which AGN feedback can be deposited isotropically into the ICM. As these ripples have a very low X-ray surface brightness contrast, it is very challenging to detect them, and they can only be seen in deep Chandra observations of nearby clusters (\citealt{Graham2008}).

We perform a set of simplified simulations to understand the detectability of sound waves in clusters using the GGM filtering technique. We create a three dimensional simulation of the Perseus cluster, modelling its density profile as a beta model with best fit parameters taken from \citet{Graham2008}, namely $\beta=0.5$, $r_{c}=30$ kpc. Into this 3D simulation we introduce sound waves with the same properties as those found in \citet{Fabian2006}, namely with a wavelength $\lambda=15$ kpc, and a constant fractional surface brightness amplitude, $h$ of 5 percent. These sound waves are assumed to be perfectly spherically symmetrical, and to be centred on the cluster core. 

We collapse the 3D simulations along the z-axis to produce projected images which we can compare to real observations. In Fig. \ref{Soundwavefig}, we shown in the top left panel a 100ks Chandra simulation of a Perseus-like cluster into which we have introduced these ripples, divided by the best fit beta model. 

It is challenging to see ripples in this image, and plotting the surface brightness residuals against radius in the bottom left panel of Fig. \ref{Soundwavefig} (red profile) shows that they have a contrast of 1-2 percent. In the top right panel we show the residuals from the GGM filtered version of the same 100 ks simulated observation, in which the ripple structure can clearly be seen. The profile of the GGM residuals is plotted in the bottom left panel (black profile) and compared to the raw surface brightness residuals. We see that the contrast of the ripples in the GGM image is significantly enhanced compared to the original image (by around a factor of 8).

Next we determined how the detectability of these sound waves decreases with increasing redshift, as the angular extent and surface brightness of the ripples decreases. We find that the minimum observing time needed to observe ripples with the same characteristics as those in the Perseus cluster increases dramatically with redshift, and this is plotted in the bottom right panel of Fig. \ref{Soundwavefig}. We find that the highest redshift at which such features can still be resolved is redshift 0.1, and that this requires observations of at least 1.5 Ms. When determining the minimum exposure time needed, we used as our detectability criterion that at least two wavelengths of the wave could be resolved by eye in the GGM filtered image. 

Due to projection effects in real images, and deviations from spherical symmetry, our values for the minimum exposure times should be interpreted as estimates. The main focus of this exercise is to understand the rate at which the exposure time needed increases with redshift. 

\begin{figure*}
  \begin{center}
    \leavevmode
   \hbox{
   \includegraphics[width=0.6\linewidth]{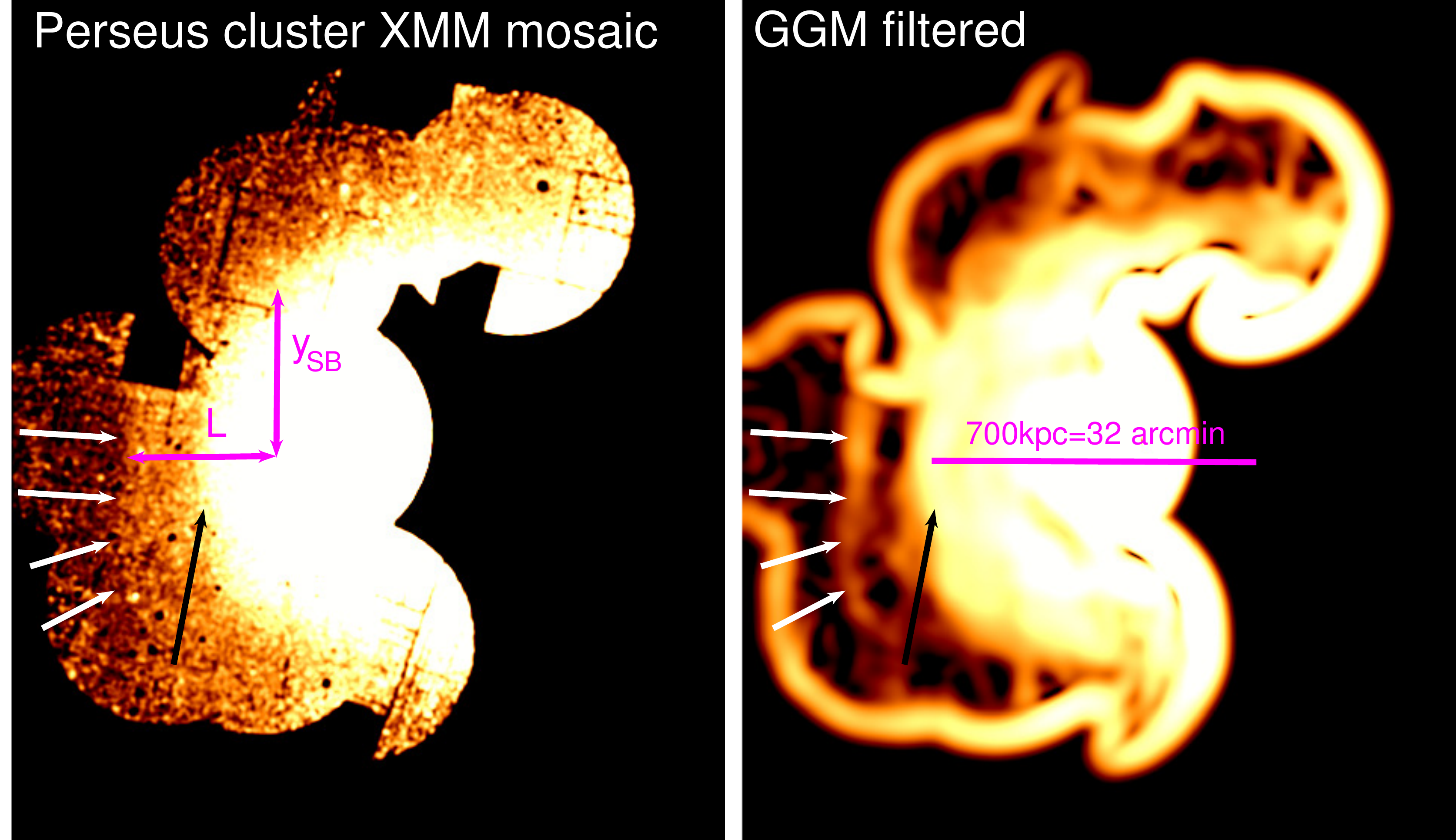}
   \includegraphics[width=0.35\linewidth]{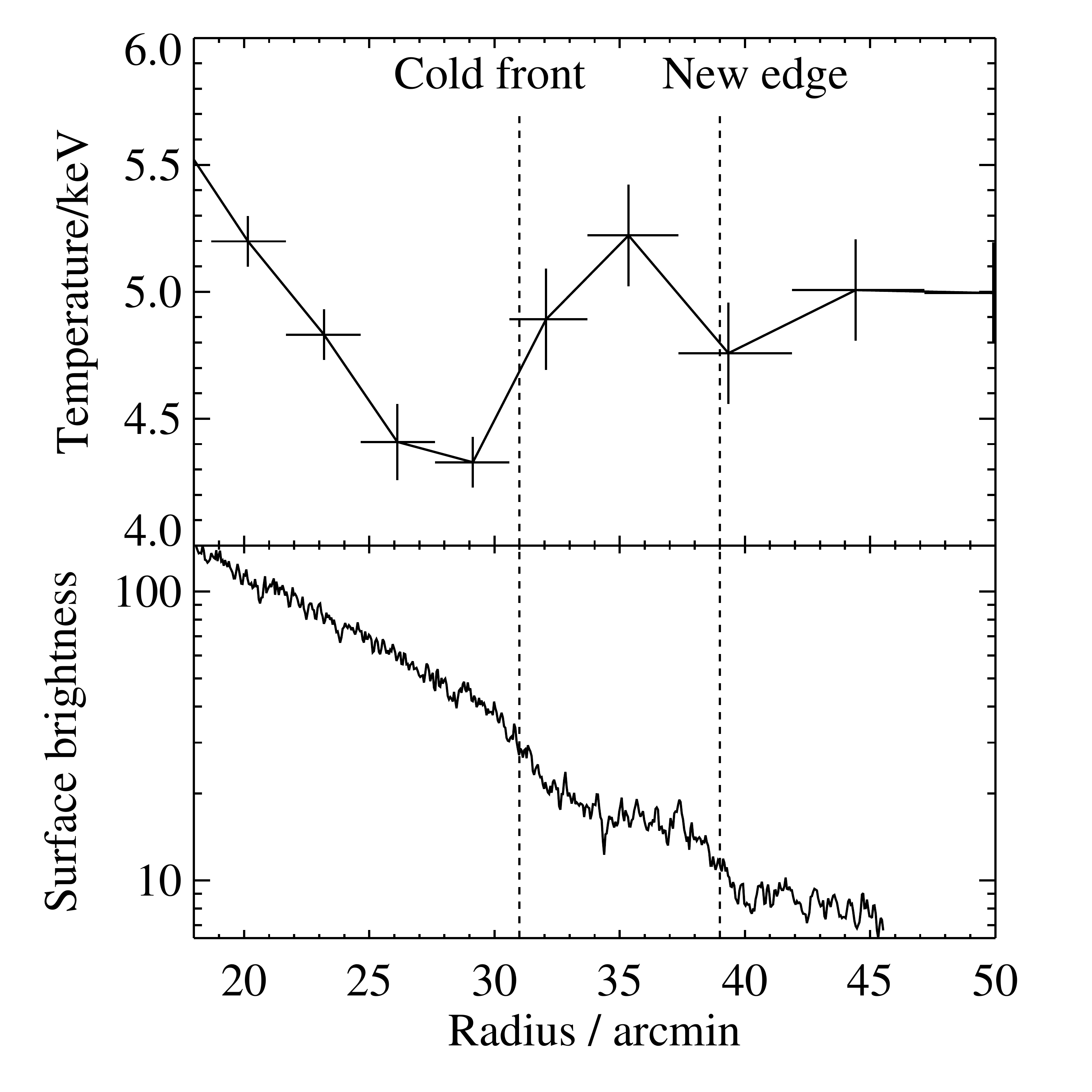}
             
                 }

      \caption{Left: The exposure corrected background and point source subtracted XMM mosaic of the large eastern cold front in the Perseus cluster. The new edge we identify is shown by the white arrows, while the cold front is shown by the black arrow. The geometric parameters $L$ and $y_{\rm SB}$ are used to estimate a Mach number if this new feature is a bow shock. Centre: GGM filtered version of the XMM mosaic, clearly showing the new edge. The purple bar shows the distance of the cold front edge from the cluster core as 700 kpc. Right:Comparing the surface brightness profile over the edges with the Suzaku temperature profile to the east from \citet{Simionescu2012}.  }
      \label{Perseus_cf}
  \end{center}
\end{figure*}

\section{Quantifying the improvement in image contrast}
\label{quantifyingimprovement}
Here we quantify the improvement in the contrast of surface brightness features afforded by the GGM filter when applied to the Chandra images of galaxy clusters. We examined a sample of deep Chandra observations of clusters (tabulated in table \ref{alldata}) which demonstrate a range of features, from merger shocks (the Bullet cluster, \citealt{Markevitch2002}, \citealt{Markevitch2006}) to cold fronts (Abell 2319), to shocks and cavities from AGN feedback (NGC 5813, \citealt{Randall2015}), to large filamentary systems in the non cool core Coma cluster.

In Fig. \ref{appendixfigure} of Appendix \ref{sec:appendix} we show the improvement in contrast achieved by running the GGM filter on these clusters. The GGM filtered images are shown in the left hand column, whilst the central column shows the original images. We find surface brightness profiles along the white rectangular strips on the images, and these are shown as the red profiles in the right hand column (dividing by the median surface brightness). The profiles start at the end of the white box marked with a white cross, and continue along its length. These surface brightness profile strips were chosen to pass through features of interest, such as the shocks in the Bullet cluster and NGC 5813, the cold front in Abell 2319. We then obtained profiles from the GGM filtered images along the same strips (normalising by the median value), and these are plotted as the black curves in the right hand column, where they can be directly compared to the simple surface brightness profiles. 

We can see that there is a dramatic improvement in the contrast of the surface brightness features when the GGM filter is used. In the Bullet cluster the contrast of the `bullet' is increased by an order of magnitude, from a factor of $\sim$3 in the raw image to a factor of $\sim$30 in the filtered image (see the region between pixel 300 and 400 in the profile plot). In NGC 5813 the contrast of the north western and south eastern shocks is improved by a factor of 3, which is similar to the level of improvement of the clarity of the central circular edge in Hercules A, and of the edges to the north east of Abell 2052. 

In PKS 1404, a possible clockwise swirling structure to the south is highlighted by the filter. In the non-cool core Coma cluster, the filter clearly shows one of the linear features to the south identified in \citet{Sanders2013}, which is extremely difficult to see in the raw data, and provides an improvement in contrast of a factor of $\sim$15. The triangular shaped edge around the two BCGs is also greatly enhanced.

\begin{figure*}
  \begin{center}
    \leavevmode
   \hbox{
   \includegraphics[width=\linewidth]{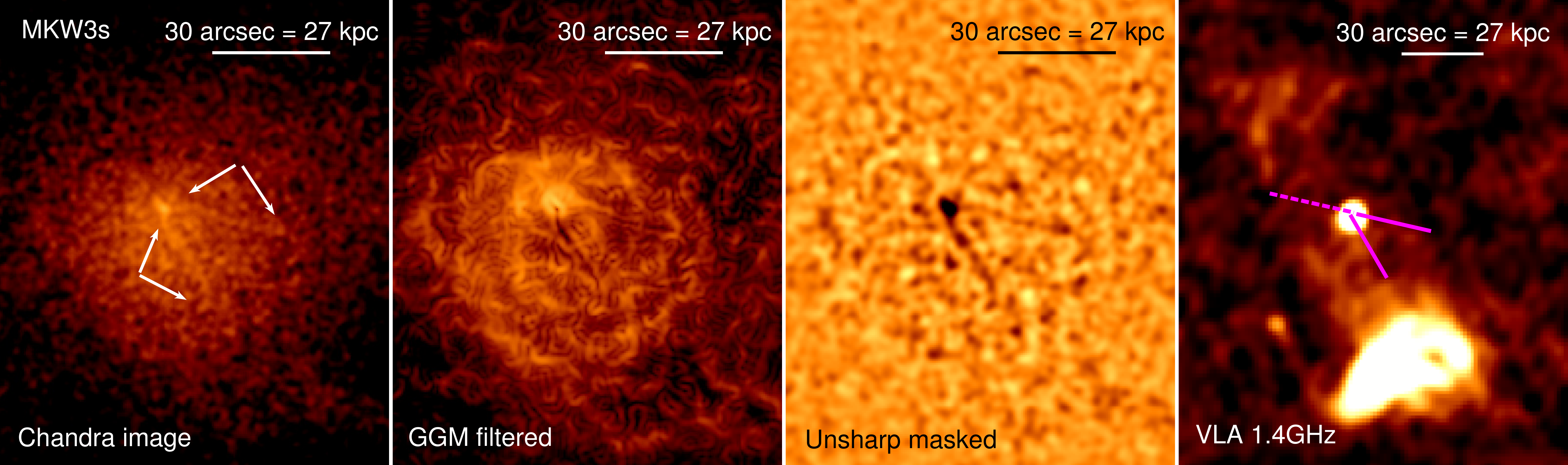}

                 }

      \caption{X-ray bright linear features in the core of MKW3s. The features can be seen in the 0.7-7.0 keV band Chandra image (leftmost panel) where they are marked by the white arrows.  They are highlighted in the GGM filtered image (second panel). They can be seen most clearly in the unsharp masked image (third panel). The rightmost panel shows a zoom out of the 1.4GHz VLA image of the radio lobes, onto which the X-ray linear features are overplotted as the solid pink lines. The dashed pink line shows that when the western feature is extended to the east it connects with the bottom of the northern radio lobe.    }
      \label{MKW3s}
  \end{center}
\end{figure*}

\section{Testing GGM filtering as a discovery tool}
\label{discoverytool}
We have begun a programme of searching for previously unidentified features in archival Chandra and XMM-Newton data using the GGM filter. Whilst the spatial resolution of XMM-Newton is significantly worse than that of Chandra, its larger field of view and collecting area allows it to map large scale structure in the outskirts of clusters, allowing us to search for large scale edges. Here we report three of the most interesting previously unidentified features we have found, providing a demonstration of the power of GGM filtered images as a discovery tool. 

\subsection{A potential bow shock at half the virial radius in Perseus?}
\label{Perseusedge}

A recent remarkable development in the study of cold fronts has been the discovery of large scale cold fronts reaching out to half the virial radius in the Perseus cluster (\citealt{Simionescu2012}, \citealt{FurushoPerseus}), A2142 (\citealt{Rossetti2013}) and RXJ2014.8-2430 (\citealt{Walker2014}). These systems feature alternating cold fronts with increasing radii on opposite sides of the cluster, suggesting that all of these features are connected by a large scale swirl. Simulations (e.g. \citealt{Ascasibar2006}) predict that as cold fronts age their geometric features should propagate outwards. However these observed large scale cold fronts lie much further out than simulations have predicted at present, so they provide a powerful test of our theoretical understanding of cold fronts. 

By applying the GGM filter to the XMM-Newton mosaic of the eastern large scale cold front in the Perseus cluster, we have identified a new edge lying 150 kpc outside the cold front, marked by the white arrows in Fig. \ref{Perseus_cf}. This new edge is roughly parallel to the cold front edge, and appears to follow the cold front's curvature to the south. Unfortunately the incompleteness of the existing XMM-Newton mosaic, combined with its uneven level of depth, prevents us from probing its behaviour to the north and south.  

This new edge is remarkably sharp. Applying the surface brightness profile fitting method we used earlier to find the widths of cold fronts, we put an upper limit on the width of the edge of $\approx$1 arcmin (22 kpc). This is similar to the Coulomb mean free path inside the edge, which is also $\approx$20 kpc, while outside the edge the mean free path increases to $\approx$40 kpc.  If this edge is part of the cold front structure, this would indicate that transport processes are still heavily suppressed at large radii, and that the mechanism causing this suppression (e.g. magnetic draping) is able to act even as the cold front rises to the outskirts. For the cold front edge itself, we obtain a upper limit on the width of 20-30 kpc, larger than the mean free path inside the cold front of 6 kpc, and of the same order as the mean free path outside it (20kpc). Unfortunately, the cold front lies near the periphery most of the XMM chips in the mosaic, further increasing the PSF blurring and limiting our ability to resolve the width of these edges, restricting us to obtaining upper limits on the widths. On axis observations with Chandra of this region are necessary to ensure that we resolve the widths of these edges.

Another possibility is that this is a bow shock, similar to the one claimed in \citet{Vikhlinin2001} in A3667. This may be possible if this large cold front in Perseus has formed from a merger like in A3667, rather than having risen from the core through gas sloshing. One immediate problem with such an interpretation however is how the cool core of Perseus could survive such a merger. Indeed, the spiral sloshing pattern observed in \citet{Simionescu2012} points towards a less energetic origin in which the geometric forms of cold fronts originating in the core have risen outwards with time to form the structure we see today.  

The observed deprojected and background subtracted surface brightness drops by a factor of 3.9 across the edge, corresponding to a density drop of a factor of $\approx$2.0. Using the Rankine-Hugoniot  shock equations, 
\begin{equation}
\frac{\rho_2}{\rho_1} = \frac{(1+\gamma) M_1^2}{2+(\gamma -1) M_1^2}
\label{eq:Mdensity}
\end{equation}
\begin{equation}
\frac{T_2}{T_1} = \frac{2 \gamma M_1^2 - \gamma + 1}{\gamma + 1} \frac{\rho_1}{\rho_2}
\label{eq:Mtemp}
\end{equation}
the Mach number corresponding to this density jump is $\approx$1.7, where $\gamma=5/3$. The temperature jump from such a shock is 1.7. This is much higher than the temperature variations observed across this feature with Suzaku in \citet{Simionescu2012} and \citet{Urban2014}, in which the temperature lies in a range 4.5-5.5 keV. A shock origin for this edge would therefore appear to be in serious tension with the observed temperature profile. Using the approach of \citet{Russell2012}, we explored whether the time taken to establish electron-ion equilibrium could account for the lack of an observed temperature jump, but found this to be an insignificant effect. 

Curiously however, the relative location of the new edge and the cold front is in reasonable agreement with a Mach $\approx$1.7 shock. As described in \citet{Vikhlinin2001}, which considered the relative position of their claimed bow shock relative to the cold front in A3667, Moeckel (1949)\footnote{\url{http://naca.central.cranfield.ac.uk/reports/1949/naca-tn-1921.pdf}}  provides a method for finding the location of the bow shock in front of an axially symmetric body. The ratio $L/y_{\rm SB}$ (where $L$ and $y_{\rm SB}$ are defined in the left hand panel of Fig. \ref{Perseus_cf}), decreases with increasing Mach number. The observed ratio for the new edge in Perseus is $L/y_{\rm SB} \approx 1$, which from Moeckel (1949) corresponds to a Mach number of $\approx 1.7$. It is entirely possible, however, that this agreement is just a coincidence. It should also be stressed that the relative locations of the shocks and cold fronts in merging clusters such as the Bullet cluster (\citealt{Markevitch2002}) and A2146 (\citealt{Russell2012}) do not agree with the Moeckel (1949) formulation, while the presence of the bow shock in A3667 has been questioned by studies of deeper Chandra observations by \citet{Datta2014}.  

An alternate possibility is that the cold front structure has changed significantly as it has risen outwards, resulting in two edges. Simulations of sloshing cold fronts have at present focussed on their behaviour in cluster cores, where the temperature increases with increasing radius. Our theoretical understanding of how cold fronts behave as they propagate out to the outskirts, where the temperature declines with radius, remains limited. If the cold front has risen outwards to 700 kpc, it will be significantly older than cold fronts seen in cluster cores, and therefore diffusion processes and instabilities will have had more time to develop. Magnetic draping is believed to be responsible for strongly suppressing transport processes over cold fronts in cluster cores, giving them widths much narrower than the Coulomb mean free path. However the way this magnetic draping behaves as cold fronts rise to large radius is uncertain.

\begin{figure*}
  \begin{center}
    \leavevmode
    \hbox{
    \includegraphics[width=0.95\linewidth]{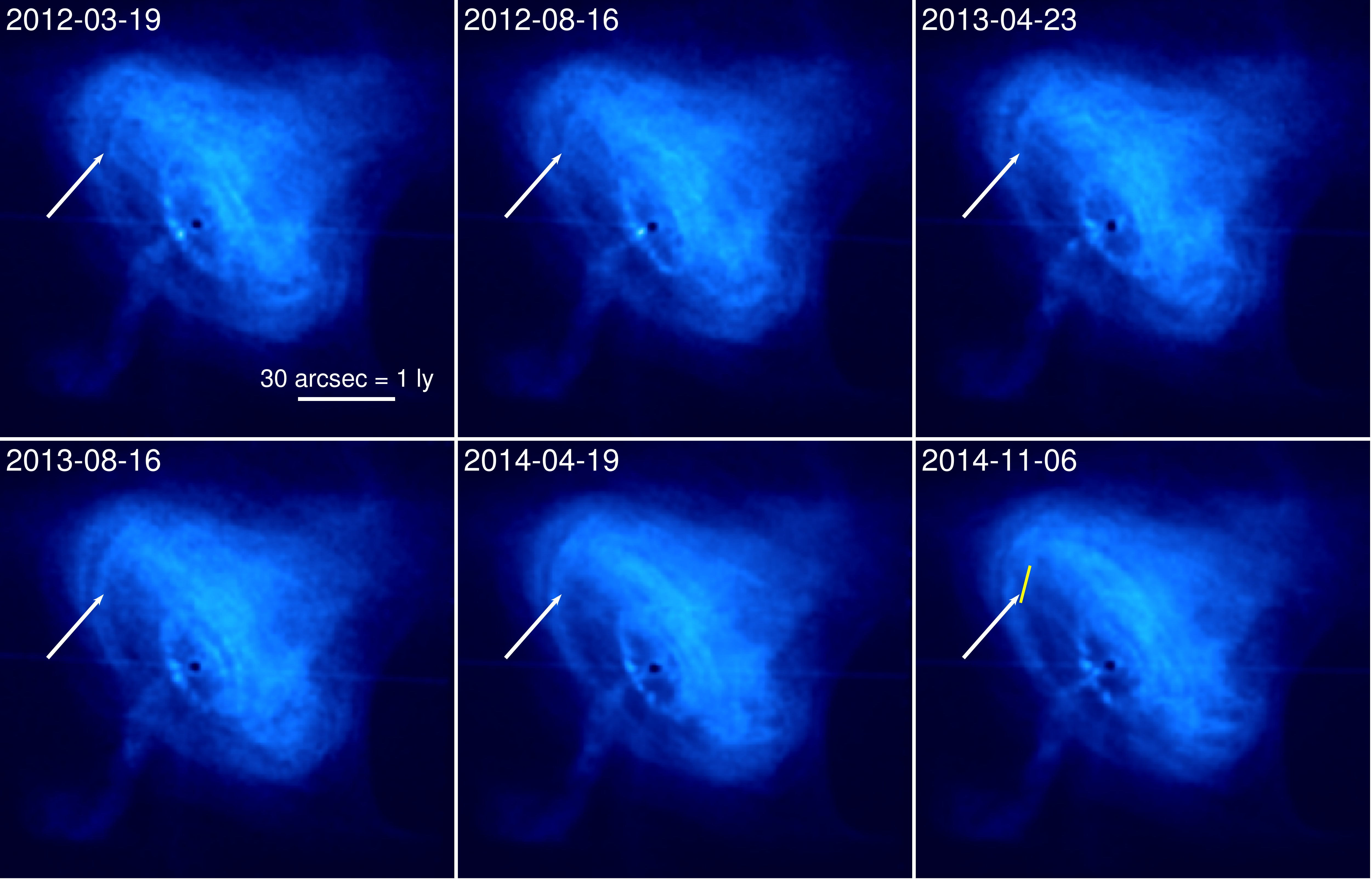}
    }
                                   \vspace{0.4cm}
             \hbox{
                 \includegraphics[width=0.95\linewidth]{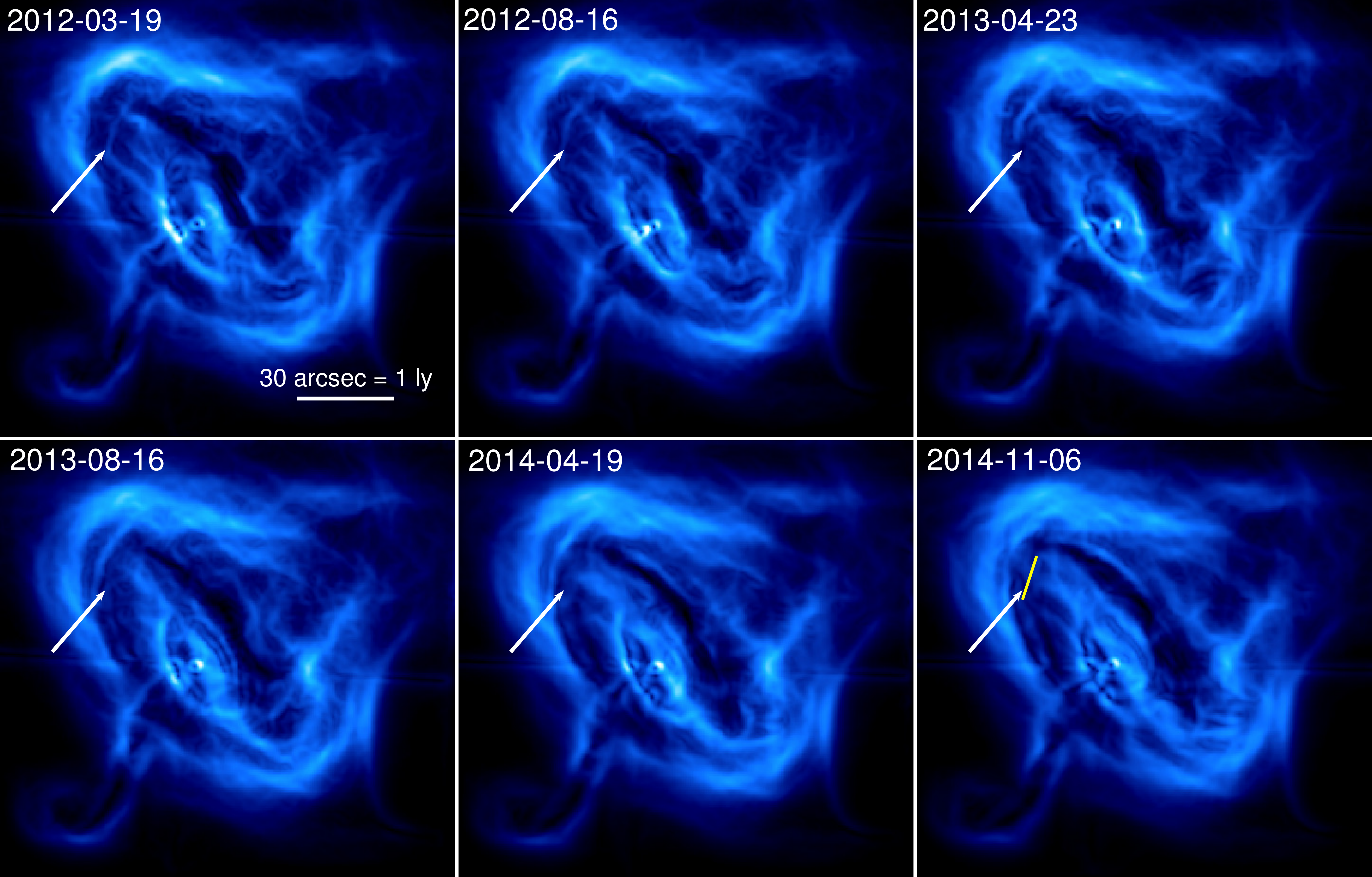} 
 }   
      \caption{\emph{The top six panels} show selected Chandra images of the Crab spaced uniformly in time from 2012-03-19 to 2014-11-06. The surface brightness edge to the north east is highlighted by the white arrow as it moves outwards. In the final panel the yellow line shows the original position of the edge on 2012-03-19. \emph{The bottom six panels} show the GGM-filtered images of the same Chandra data, in which the north eastern edge can be more clearly seen as it moves outwards. All of the images have had their coordinates matched.}
      \label{Crab_images}
  \end{center}
\end{figure*}

\subsection{Linear features in MKW 3s}
\label{MKW3ssection}
Using the GGM filter we have identified two previously unseen, unusual linear features in the core of the cluster MKW 3s, shown in Fig. \ref{MKW3s}. Previous studies of this cluster (\citealt{Mazzotta2002}) focussed on the larger scale ICM distribution. The two features are each around 30 kpc long with a uniform FWHM of $\sim$ 2 kpc, extend radially outwards from the BCG to the south west, and have an angle between them of $\approx$45 degrees. The features are most clearly seen in the unsharp masked image shown in the third panel in Fig. \ref{MKW3s}, in which we subtract a heavily smoothed image (smoothed by a Gaussian kernel with a FWHM of 20 pixels) from a lightly smoothed image (FWHM of 4 pixels). 

In the rightmost panel of Fig. \ref{MKW3s} we zoom out slightly to show the 1.4GHz VLA data, which features two prominent radio lobes, with the locations of the X-ray linear features overplotted as the solid pink lines. We see that the southern linear feature is aligned roughly along the expected jet axis for the radio lobes. The angle subtended by the southern radio lobe from the BCG location is the same as the angle between the linear features (45 degrees), however they are not aligned. If we extend the western linear feature to the east (shown by the dashed pink line), we see that it connects with the `finger' of radio emission extending down from the northern radio lobe. 

These correlations between the X-ray linear features and the radio lobes point towards a jet origin for the former. One possibility is that the jet direction has precessed by $\sim$45 degrees, and that the X-ray bright linear features are the rims of the cone formed by the jet as it has precessed. The fact that the southern radio lobe is much brighter than the northern one suggests that the southern jet has a significant component directed towards us, while the northern jet could be pointed away from us. This line of sight projection effect could explain why only X-ray linear features associated with the southern jet can be seen. 

We attempted to examine the spectrum of the excess linear features, using area scaled local background regions to subtract the cluster contribution to the X-ray emission. Unfortunately the number of excess counts is relatively low ($\approx$ 500), and insufficient to allow a meaningful spectral analysis.

The identification of these previously unseen features in MKW 3s highlights the power of the GGM filter when applied to large samples of Chandra images. Because the GGM filter allows the gradient structure on multiple scales to all be mapped onto just one image, it dramatically increases the ease with which features can be identified, without any need to `tailor' the filter to maximise the contrast in images.  Whilst it is possible to make a clearer image using unsharp masking, this required considerable iteration and checking to maximise the contrast of the linear features. If we had blindly applied unsharp masking to MKW 3s, without knowing what we were looking for, we would almost certainly have missed these features.

\subsection{The Crab Nebula}

Extending our search beyond galaxy clusters, the Crab nebula provides an excellent target for searching for variations in X-ray gradients on time-scales of months to years. Detailed optical and X-ray images (\citealt{Hester1995}, \citealt{Weisskopf2000}, \citealt{Hester2002}) have revealed a great deal of structure in the inner regions of the nebula. There is an inner ring (believed to be a termination shock from the relativistic wind of particles accelerated by the central pulsar), toroidal structure, and two oppositely directed jets from the pulsar (for a review see \citealt{Hester2008}). Its exceptional X-ray brightness allows detailed X-ray images of the structure surrounding the pulsar to be made with very brief (less than 1ks) exposures with Chandra. It is well known that the inner ring structures around the central pulsar feature rapidly moving wisps, which have been observed to move at speeds $\sim$0.5$c$ (e.g. \citealt{Hester2002}). We analysed the archival Chandra ACIS-S data (tabulated in table \ref{alldata}) on the Crab using the GGM-filter and searched for signs of coherent motions in the X-ray gradients. 

\begin{figure}
  \begin{center}
    \leavevmode
    \includegraphics[width=0.95\linewidth]{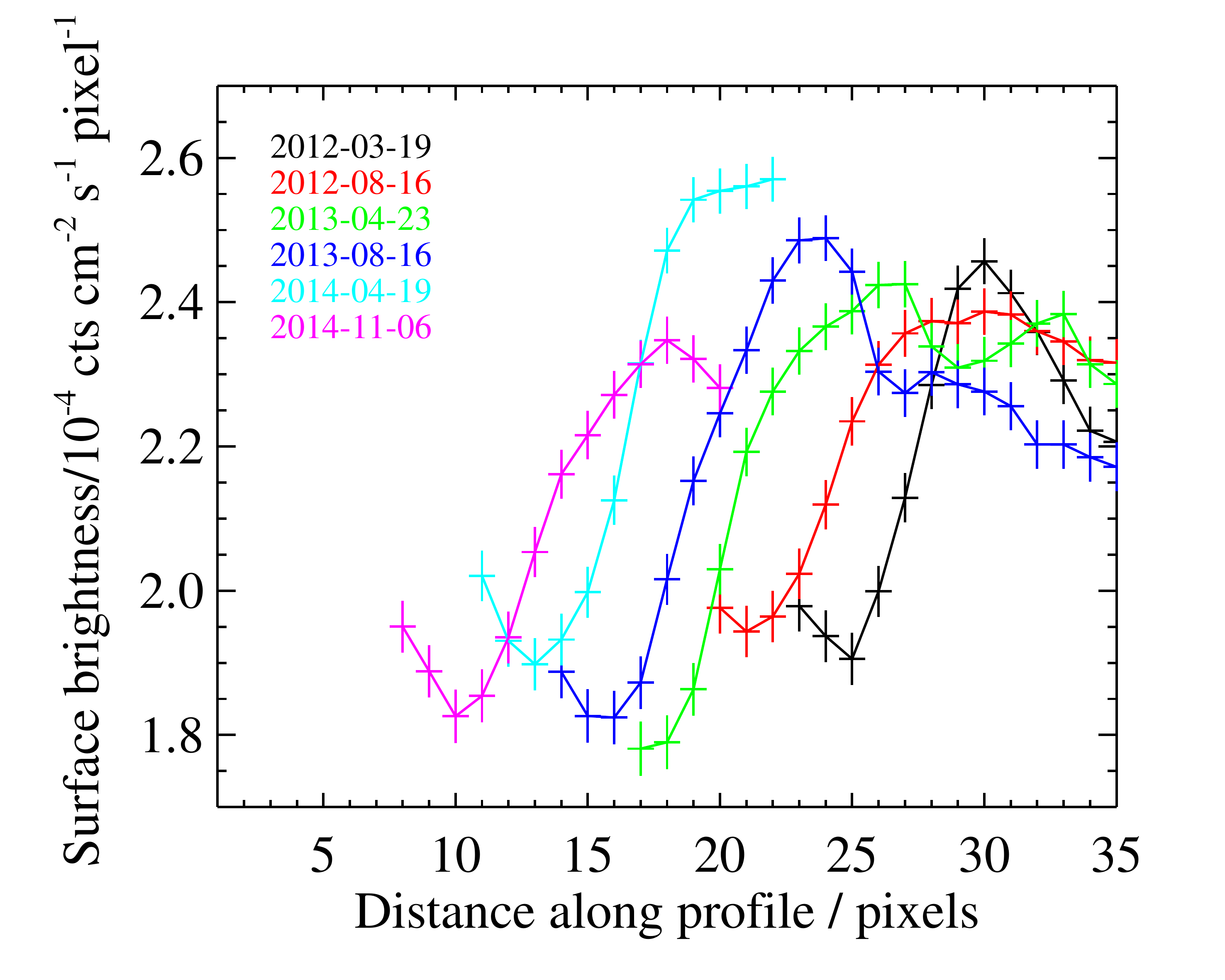}            
      \caption{The evolution of the surface brightness profile over the outwardly moving edge in the Crab nebula for the observations we study.}
      \label{Crab_profiles}
  \end{center}
\end{figure}

We identified one previously unremarked upon feature, shown in the sequence of images in Fig. \ref{Crab_images}. An X-ray edge to the north east of the Crab can clearly be seen in data starting from 2012-03-19, labelled with the white arrow in Fig. \ref{Crab_images} which shows the Chandra images (top six panels) and the GGM-filtered images (bottom six panels). This X-ray edge grows and moves outwards, and is much further from the central pulsar (a projected distance of 1.3 ly) than previously identified motions. In Fig. \ref{Crab_profiles} we show how the surface brightness profile over the edge evolves as it moves outwards. In the most recent observations available, from 2014-11-06, the edge has moved by 0.25 light-years in the plane of the sky from its original location on 2012-03-19. The original location on 2012-03-19 is shown by the yellow line in the images from 2014-11-06. We can therefore put a lower limit of 0.1$c$ on the velocity of this feature.

Using all of the Chandra observations available from 2010 to the present, we have constructed a movie showing the rise and growth of this X-ray edge, which can be accessed here\footnote{\url{http://www-xray.ast.cam.ac.uk/~swalker/Crab\_movie.mp4}}.

\section{Conclusions}

We have explored applications of the Gaussian Gradient Magnitude (GGM) filtering technique for Chandra data, focussing on cold fronts in galaxy clusters. We find that this technique is able to resolve substructures behind the cold fronts in Abell 2142 and Abell 496, which are qualitatively consistent with the structures expected from the onset of Kelvin Helmholtz instabilities. 

We perform tests and find that the GGM method provides a straightforward way of mapping variations in the gradient along the length of cold fronts in systems where the density jump ratio is relatively constant. This allows the variations in the widths of cold fronts to be mapped far more easily than present methods involving dividing the cold front into multiple sectors and fitting the surface brightness profiles with models.     

To quantify the improvement in the contrast of X-ray surface brightness edges in galaxy clusters, we tested the filter on a range of deep Chandra observations of clusters demonstrating different features, such as shocks, cold fronts, filaments and cavities. We find that the contrast of edges is improved by at least a factor of 3 in all of the observations we examine, and that the improvement can be as a high as a factor of $\sim$15 (e.g. in the Coma cluster). The filter provides a straightforward way of identifying spiral structure in clusters (e.g. PKS 1404), and of enhancing subtle variations in extended emission, such as the southern filament in the Coma cluster. Our simulations of sound wave ripples have shown that their contrast is enhanced significantly, by a factor of $\approx$8.

We have demonstrated that the GGM filter provides a powerful tool for discovering previously unseen features in archival Chandra and XMM-Newton observations. Its power stems from the fact that it is able to map surface brightness edges on multiple scales, and combine these to produce a single image. This is a significant advance over unsharp masking techniques, which are only able to improve the contrast of features on a single spatial scale, and have to be finely tuned to the specific image being considered. The GGM filter's versatility allows it to be applied to data with little initial configuration, and is well suited to an automated pipeline analysis of the vast Chandra and XMM-Newton archives.

In the outskirts of the Perseus cluster we have identified a new edge in the XMM-Newton mosaic, lying 850 kpc from the cluster core to the east and around 150 kpc outside a known large scale cold front. One possibility is that his is a bow shock, with a Mach number of $\approx$1.7 inferred from the density jump, which is agrees with the geometric location of the feature relative to the cold front. However such a shock should be accompanied by a temperature jump of a factor of 1.7, which is not observed. It is challenging to see how such a shock could exist without the associated merging event disrupting the cool core of Perseus. This interpretation is also at odds with the observed large scale spiral sloshing pattern seen in Perseus (\citealt{Simionescu2012}) which points towards a gentler origin in which the cold fronts originate from gas sloshing in the core and rise outwards from the core with time.

We have identified unusual linear features in the core of the cluster MKW 3s, extending 30 kpc radially from the BCG to the south west in a `V' shape with an opening angle of 45 degrees, which had not been seen previously. Their opening angle is very similar to the angle extended by the southern radio lobe from the BCG, and the southern linear feature lines up well with the expected jet direction. We therefore postulate that these linear features are the result of a precessing jet from the BCG, which may be being beamed towards us to the south, and away from us to the north, explaining why we do not see a northern counterpart. It is possible that the features result from the edges of a cone traced by the precessing jets. 

We have further tested the discovery power of the GGM filter to studies of the Crab nebula, whose extended X-ray emission shows structure which varies on time-scales of months. In the Crab, the GGM filter has allowed us to identify a structure which first emerges in the observation taken on 2012-03-19 before growing and moving outwards. From 2012-03-19 to 2014-11-06 the structure moves by 0.25 ly in the plane of the sky, allowing us to place a lower limit of 0.1$c$ on the velocity of this feature. This feature lies in the outer regions of the X-ray torus structure, well outside of the central region where moving wisps have been previously observed (e.g. \citealt{Hester2002}). It is possible that this feature is a continuation of the variability seen in the central regions of the Crab nebula around the central pulsar, as it propagates outwards.

\section*{Software}

The code used in this paper is available at \url{https://github.com/jeremysanders/ggm}

\section*{Acknowledgements}
SAW and ACF acknowledge support from ERC Advanced
Grant FEEDBACK. This
work is based on observations obtained with the \emph{Chandra} observatory, a 
NASA mission and the \emph{XMM-Newton} observatory, an ESA mission.
\bibliographystyle{mn2e}
\bibliography{Edges_v2}

\appendix

\section[]{GGM filtered images}
\label{sec:appendix}

\begin{figure*}
  \begin{center}
    \leavevmode
    
     \hbox{ 
     \includegraphics[width=0.6\linewidth]{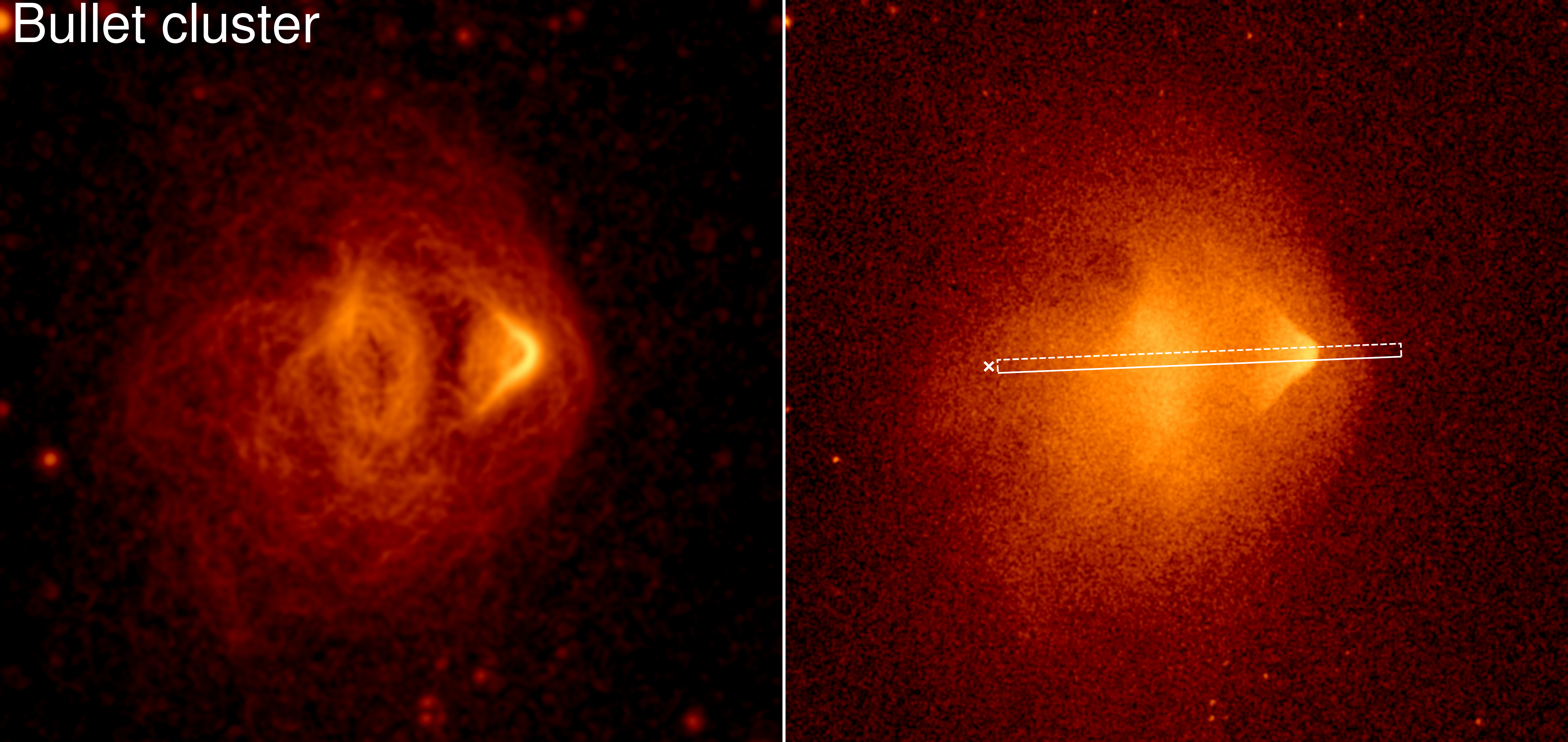}
     \includegraphics[width=0.3\linewidth]{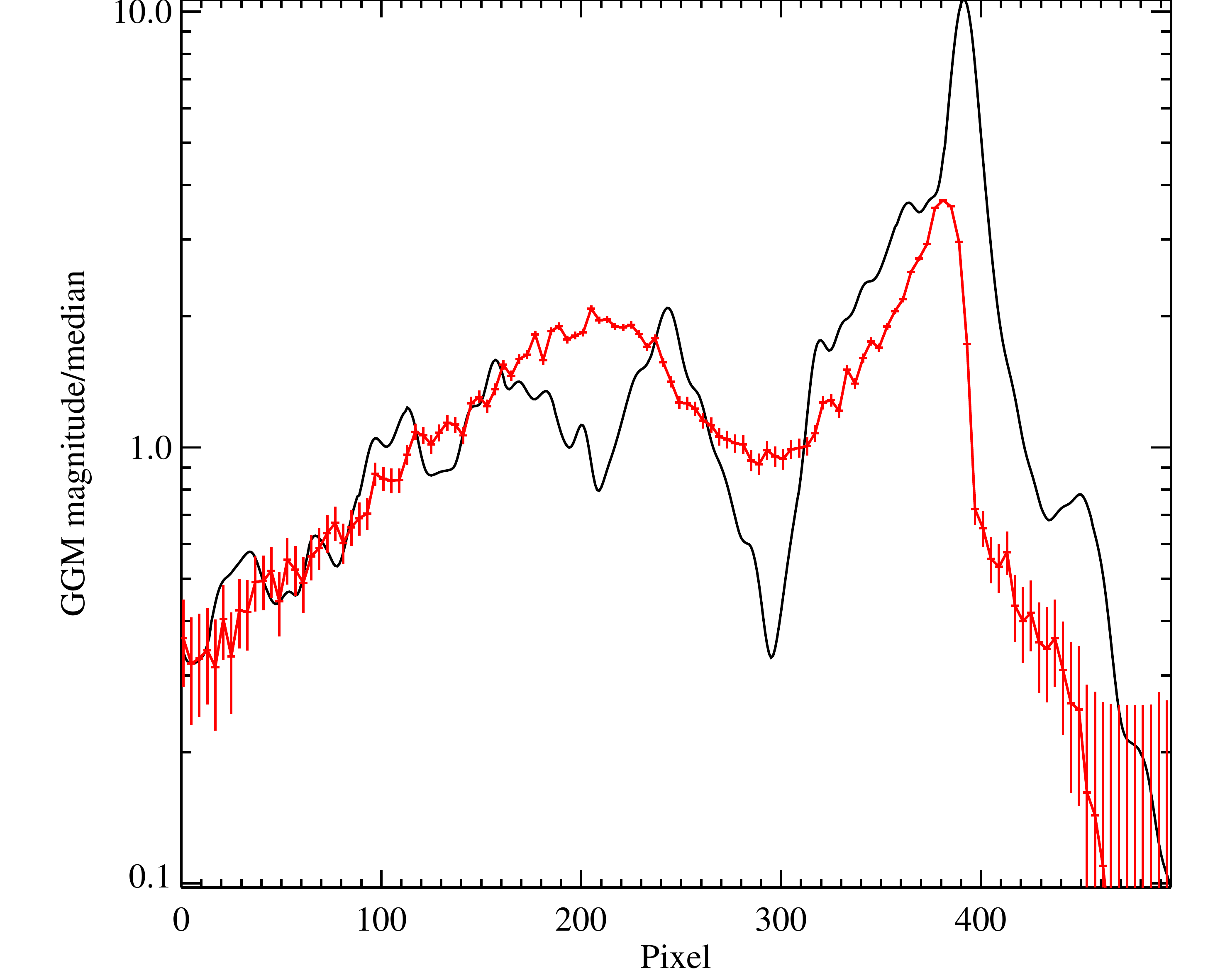}
   }
   
   \hbox{    
   \includegraphics[width=0.6\linewidth]{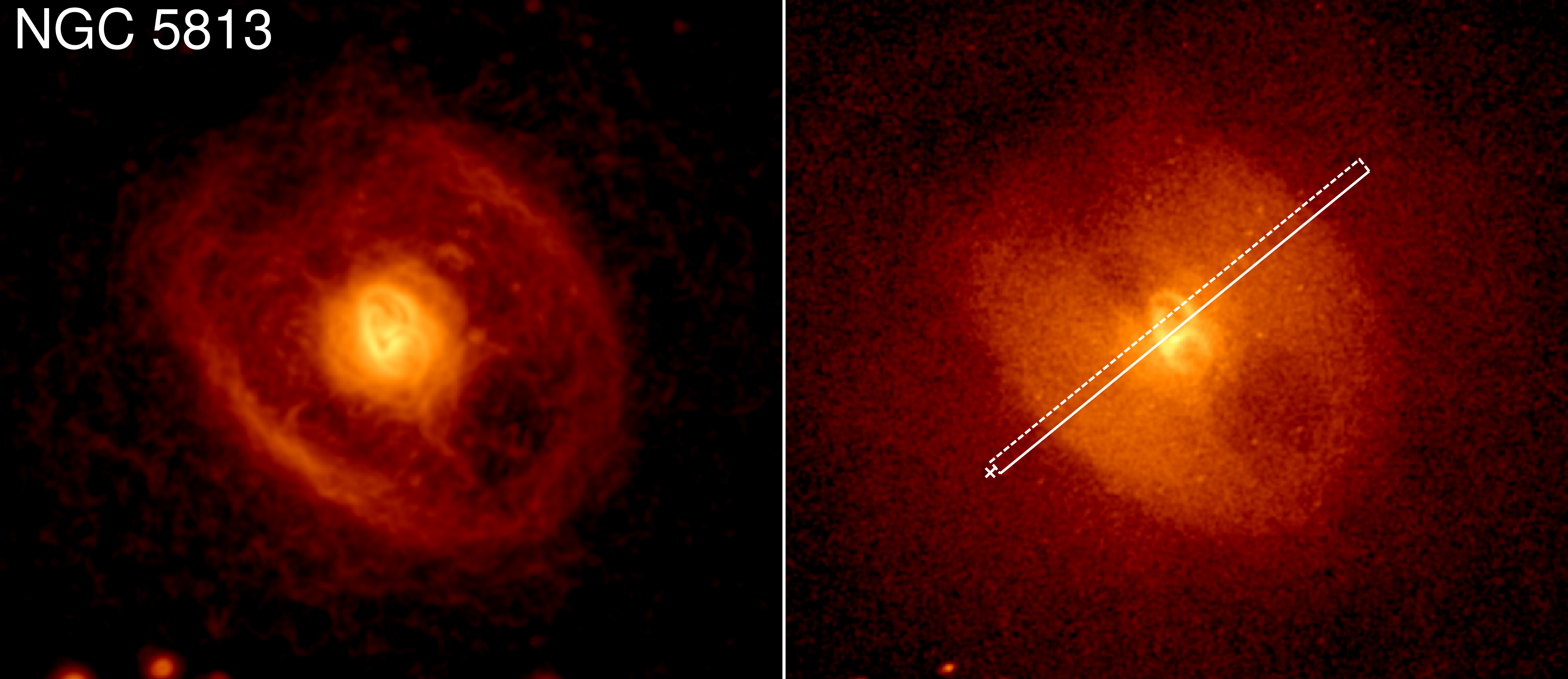}
     \includegraphics[width=0.3\linewidth]{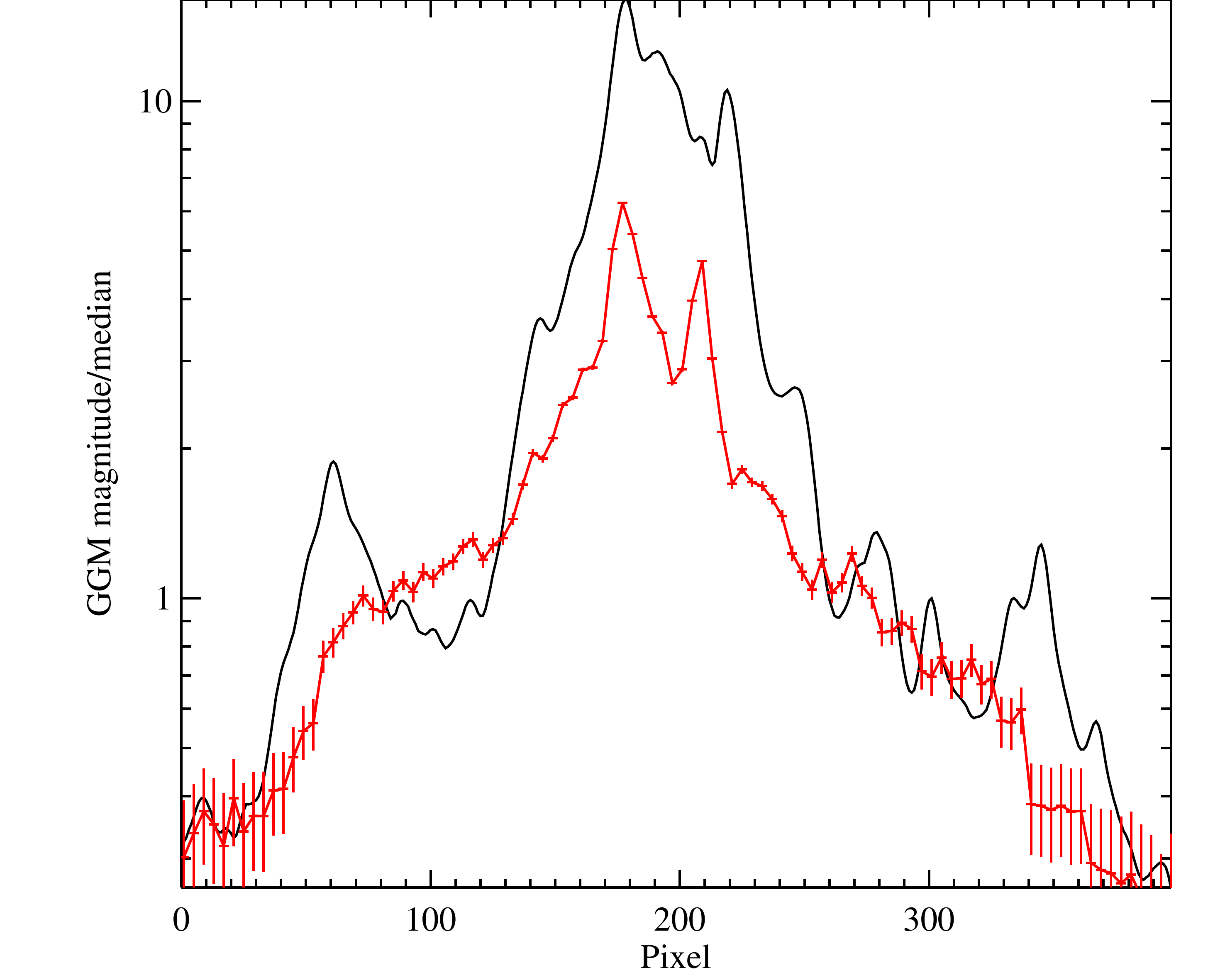}

   }
   
    \hbox{     
       \includegraphics[width=0.6\linewidth]{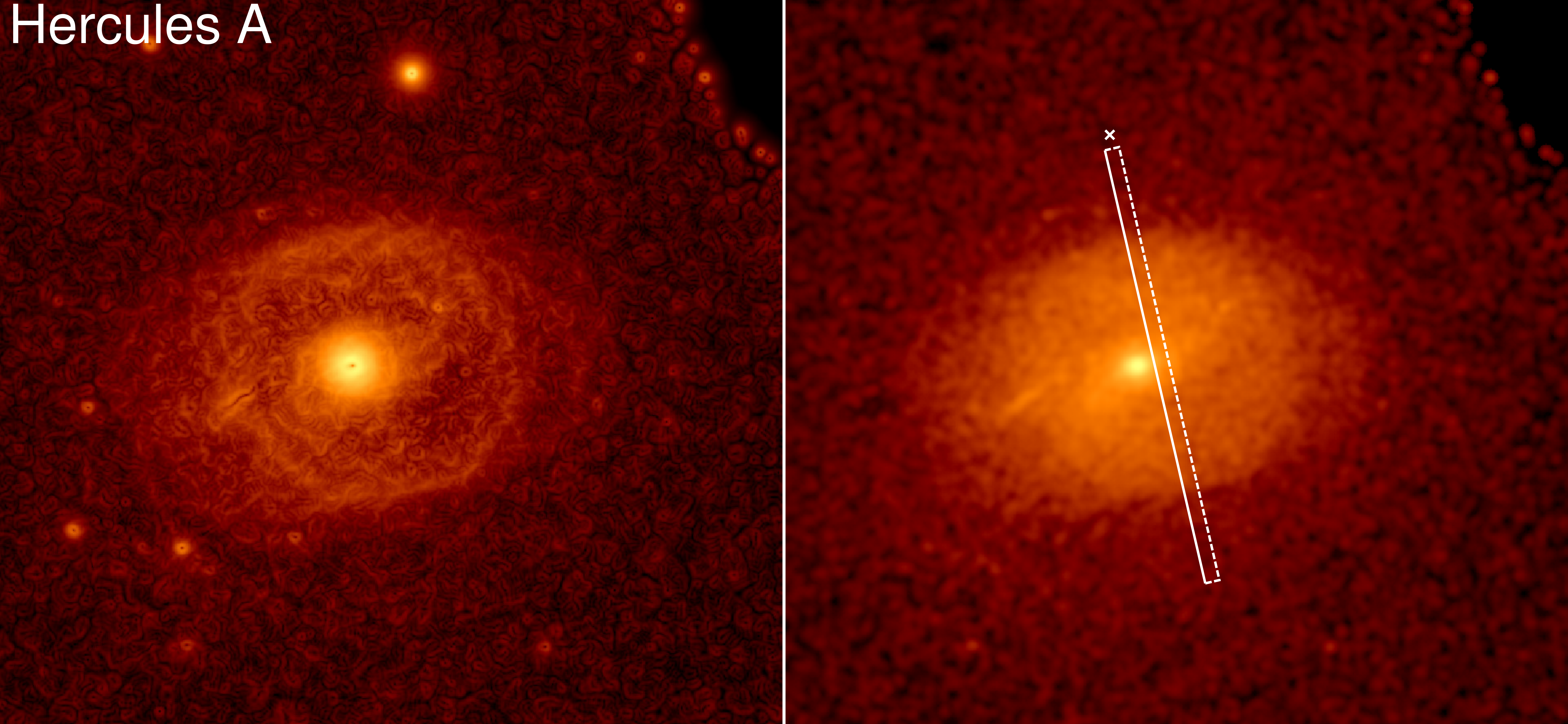}
     \includegraphics[width=0.3\linewidth]{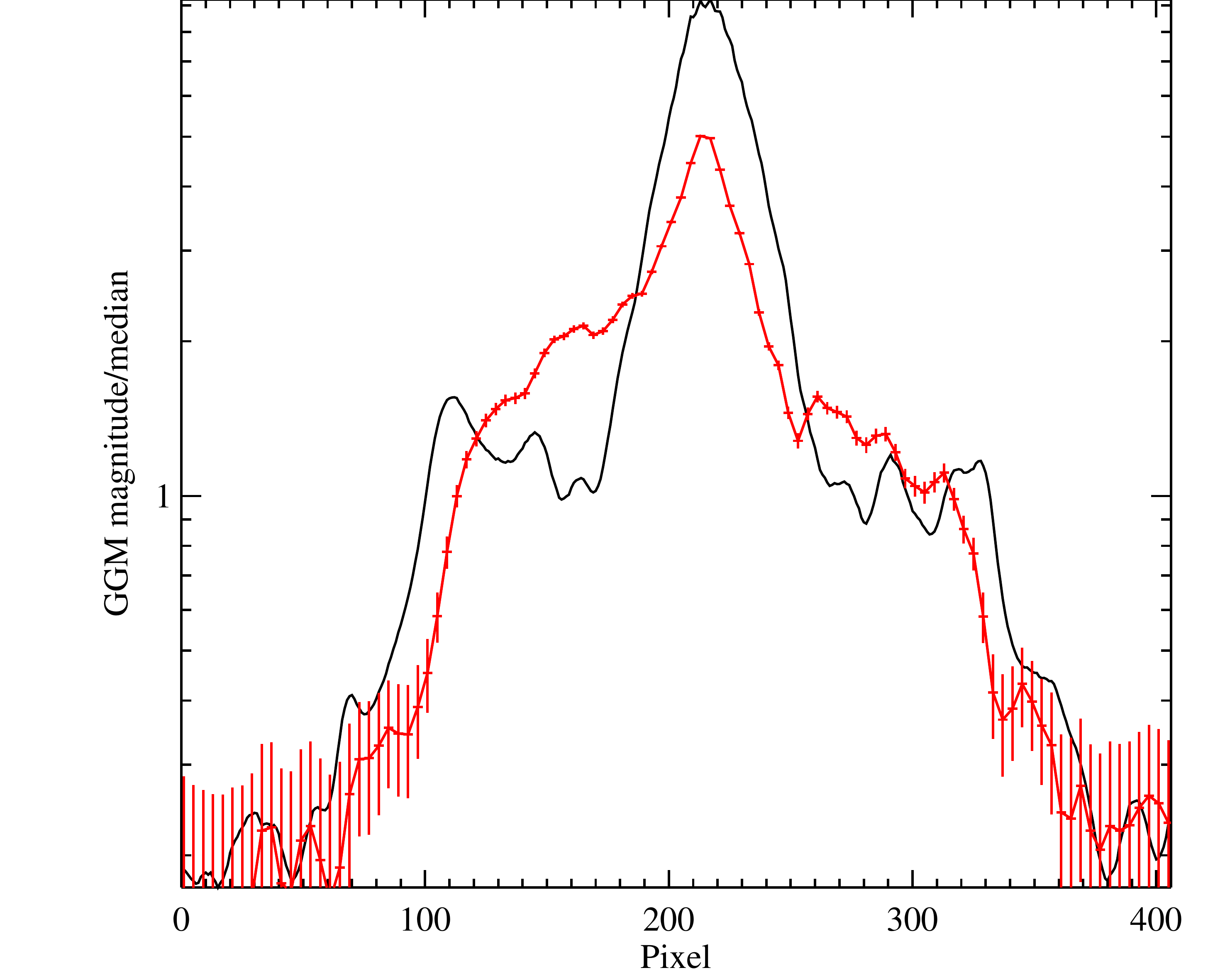}
             
   }
    \hbox{     
          \includegraphics[width=0.6\linewidth]{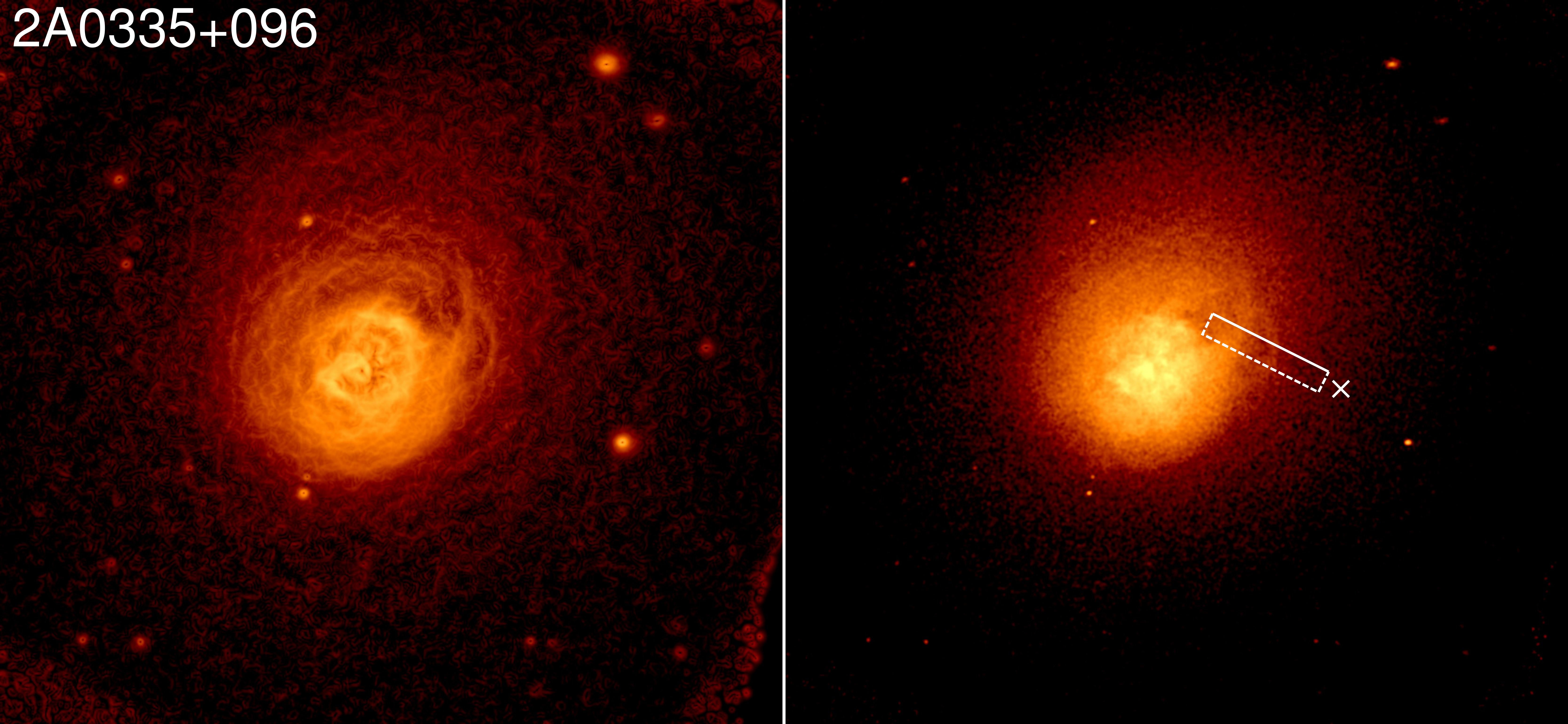}
     \includegraphics[width=0.3\linewidth]{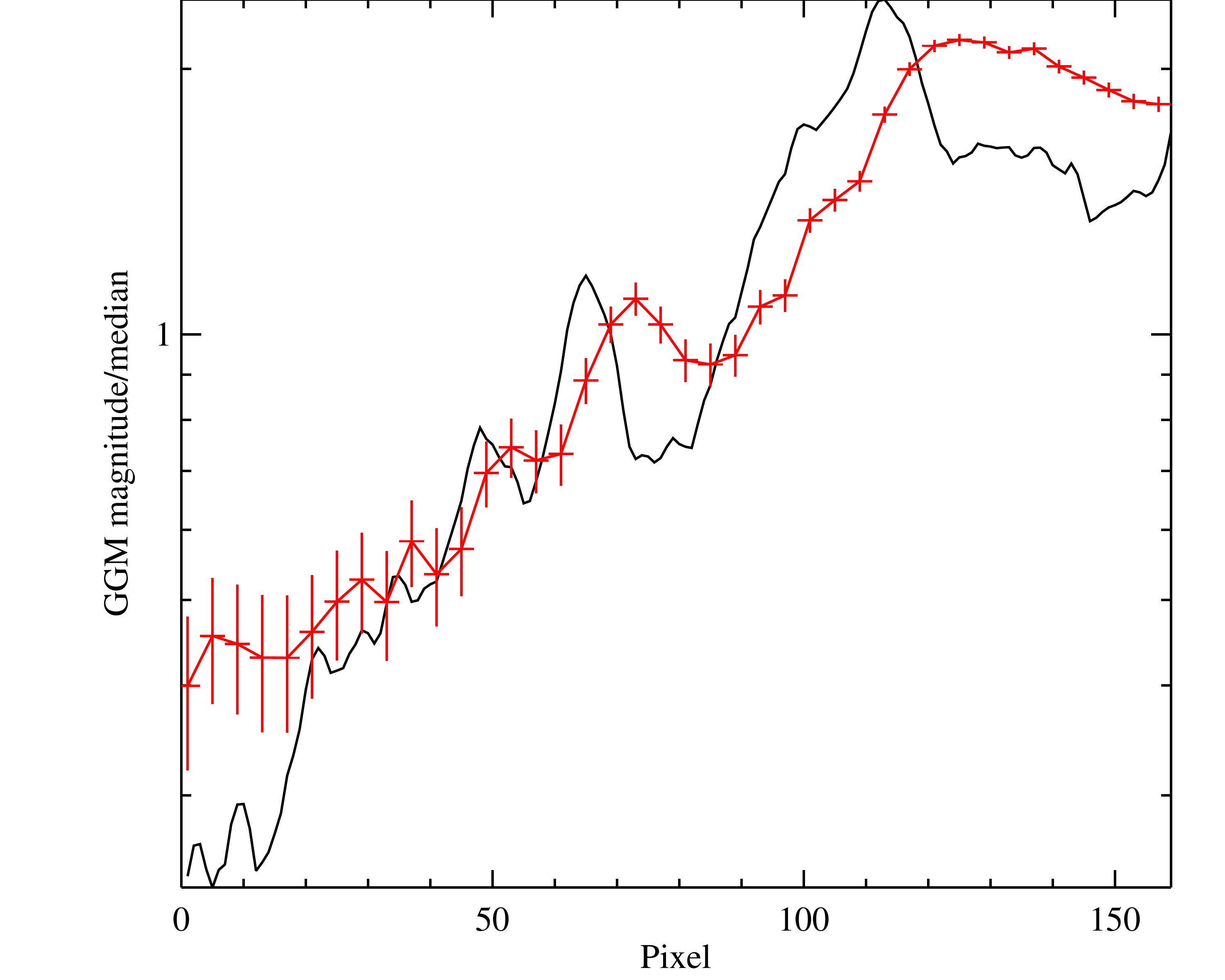}

   }
   
      \caption{Comparing the contrast of a variety of features in Chandra observations of galaxy clusters before (centre panels) and after (left panels) the application of the GGM filter. Profiles are taken along the white strips (starting at the end marked with a cross) in both images and compared in the right hand panel, where the red profile is across the original image and the black profile is across the filtered image. Both profiles have been normalised by their median to aid comparison. }
      \label{appendixfigure}
  \end{center}
\end{figure*}

\begin{figure*}
  \begin{center}
    \leavevmode
    
 \hbox{ 
     \includegraphics[width=0.6\linewidth]{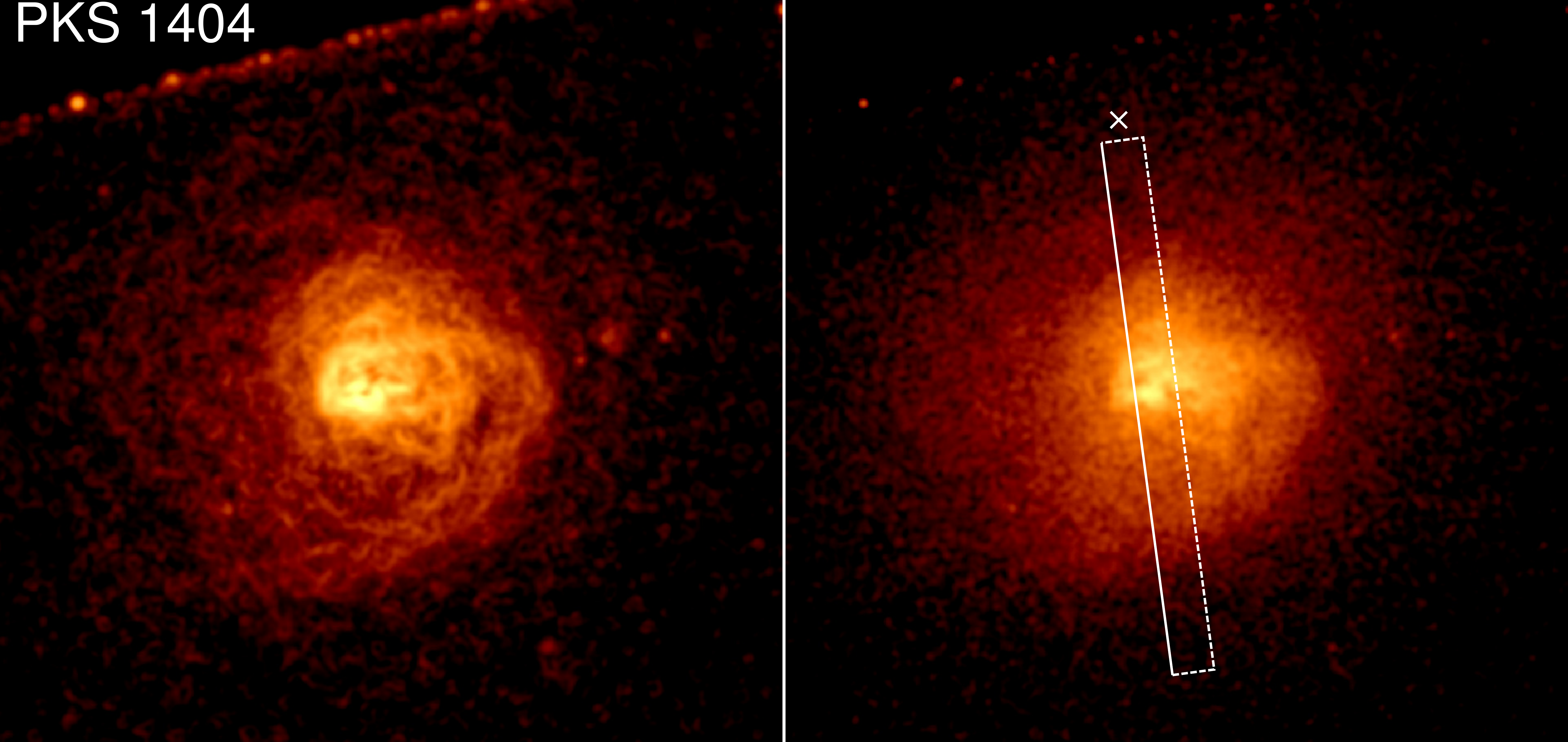}
     \includegraphics[width=0.3\linewidth]{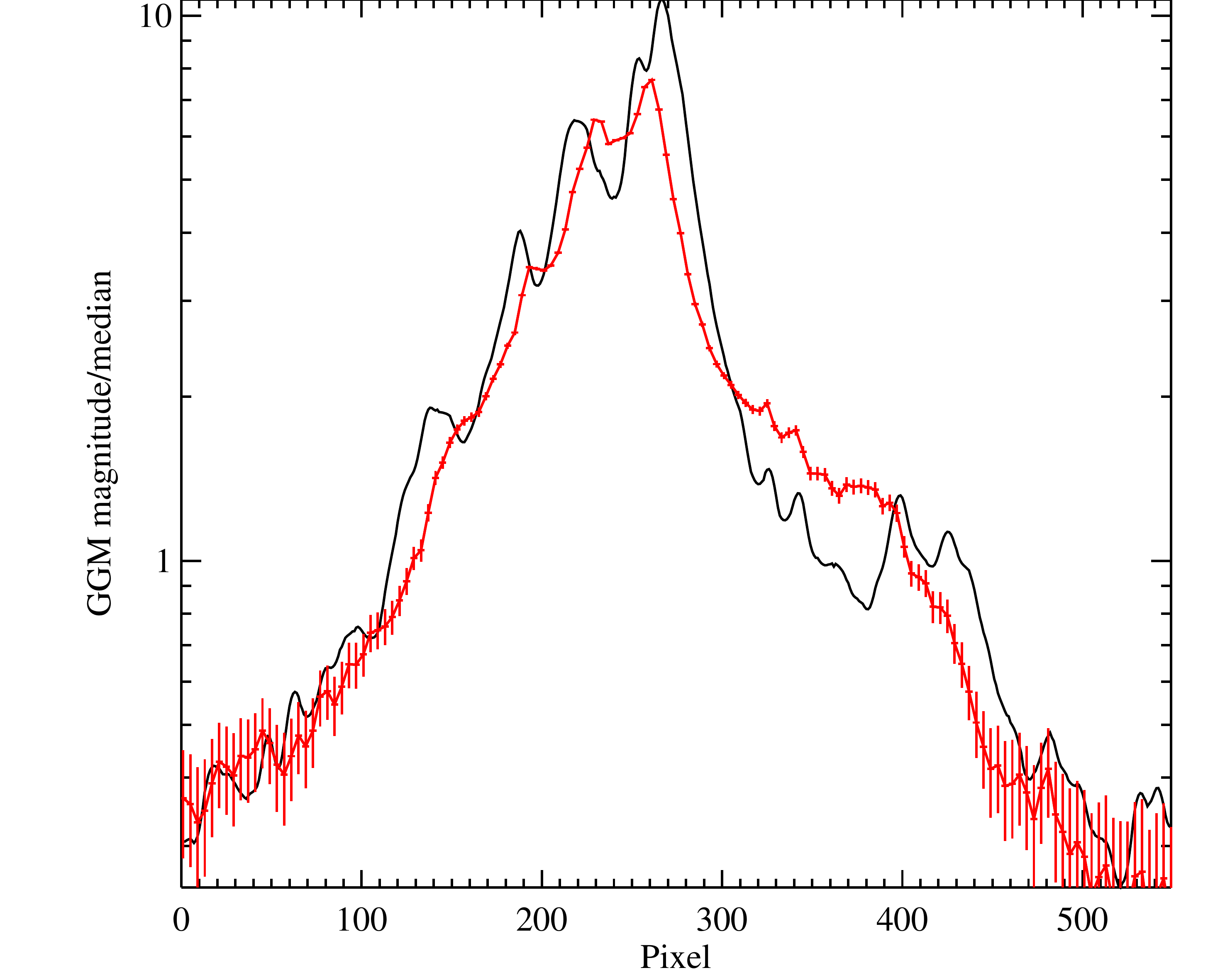}
   }
   
   \hbox{    
   \includegraphics[width=0.6\linewidth]{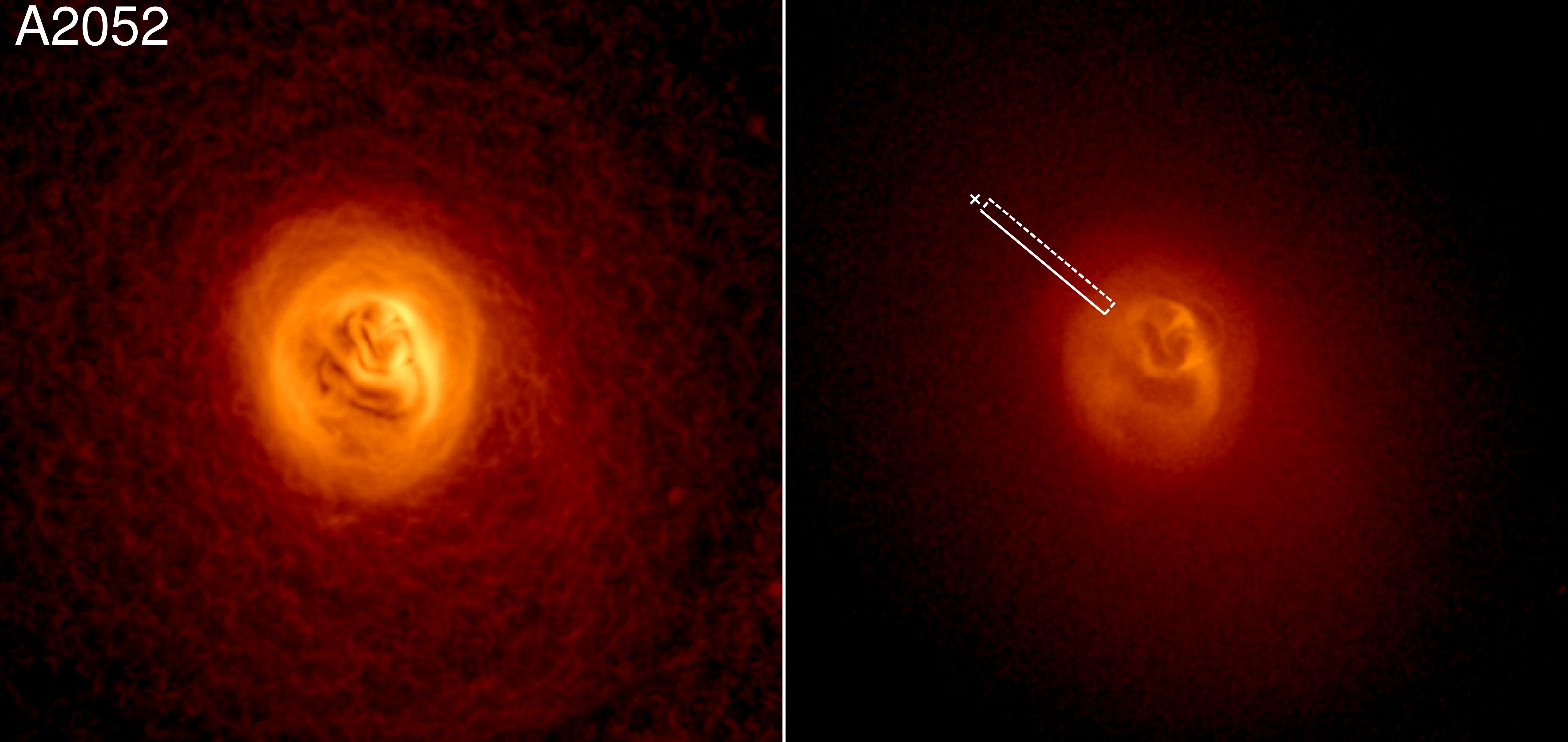}
     \includegraphics[width=0.3\linewidth]{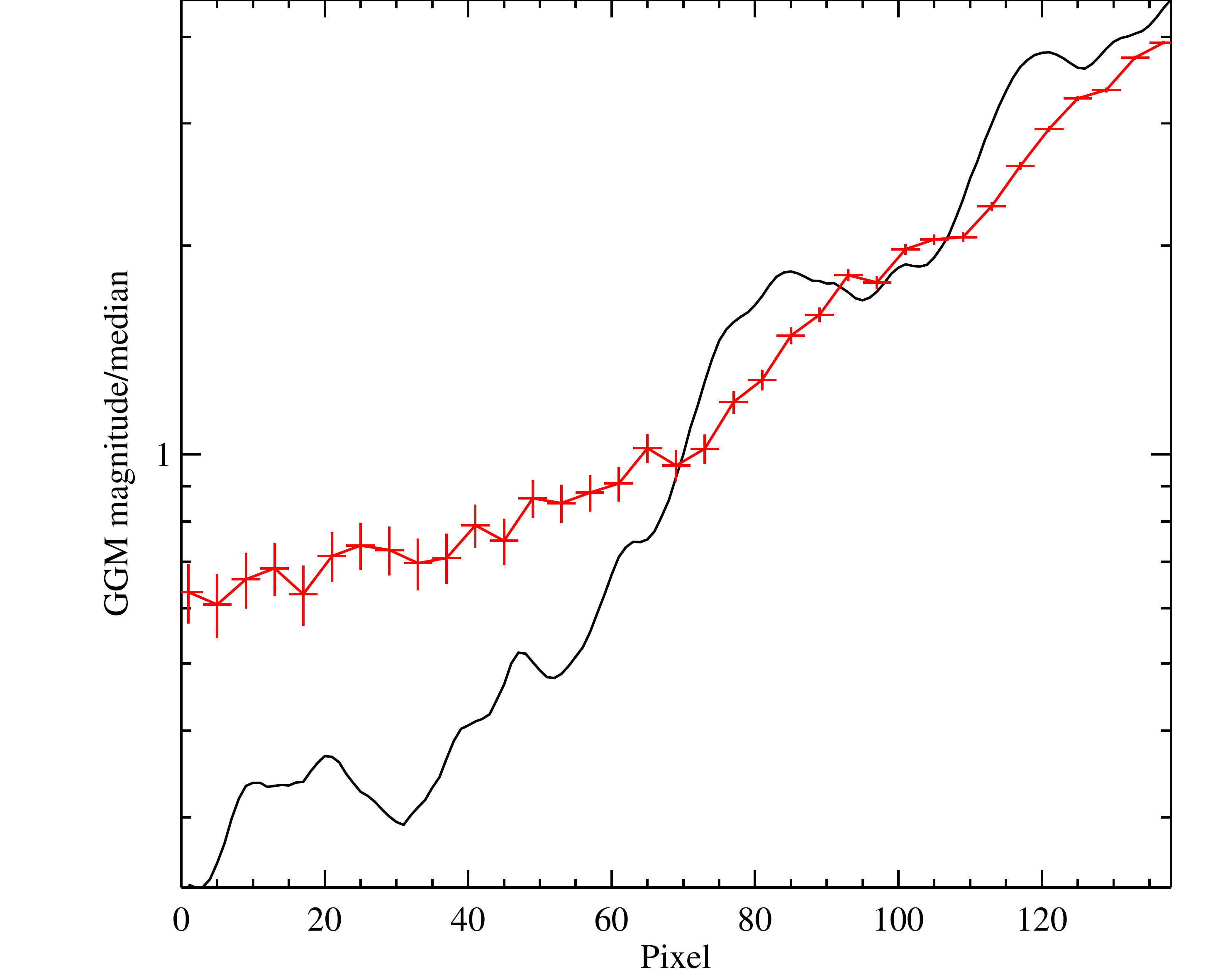}

   }
   
    \hbox{     
       \includegraphics[width=0.6\linewidth]{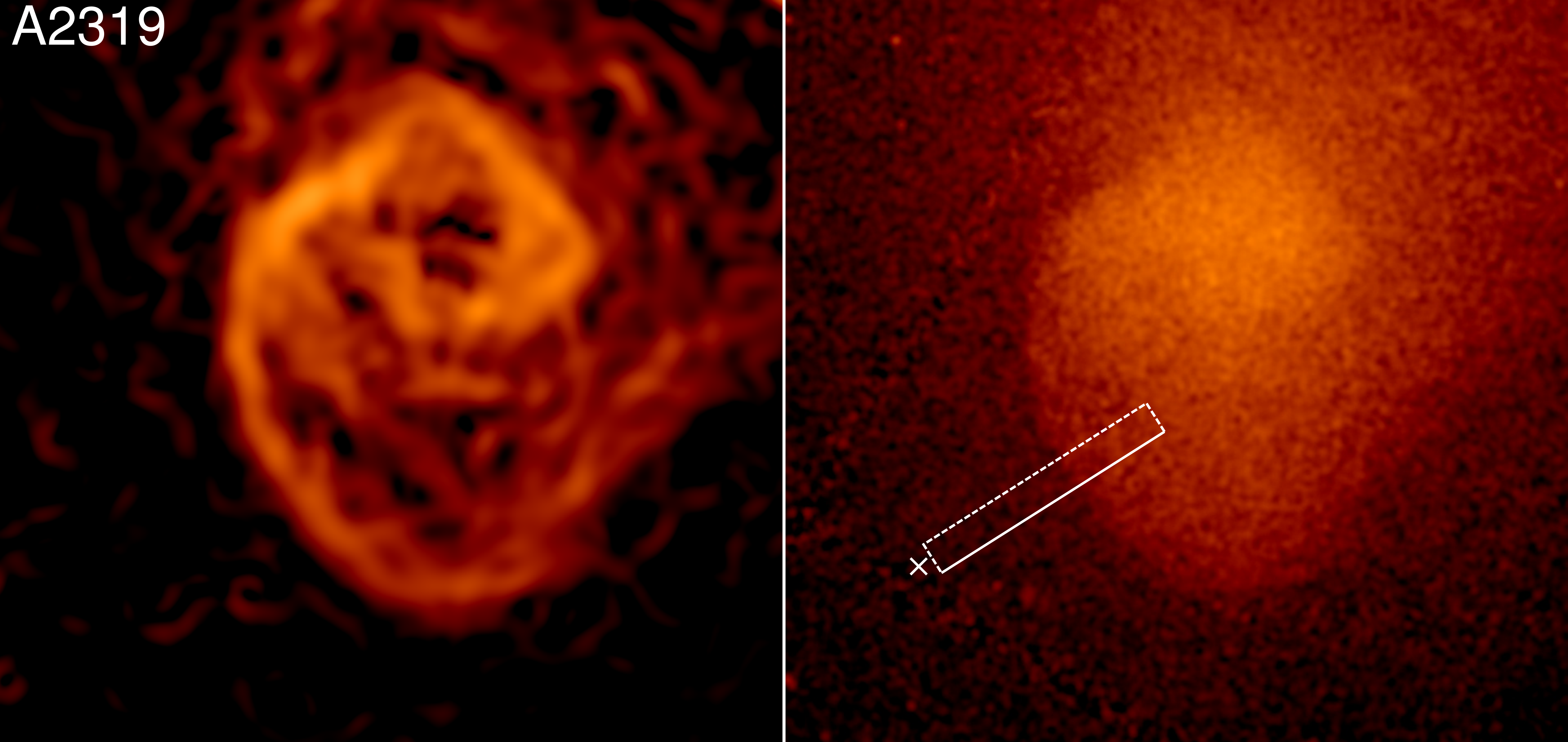}
     \includegraphics[width=0.3\linewidth]{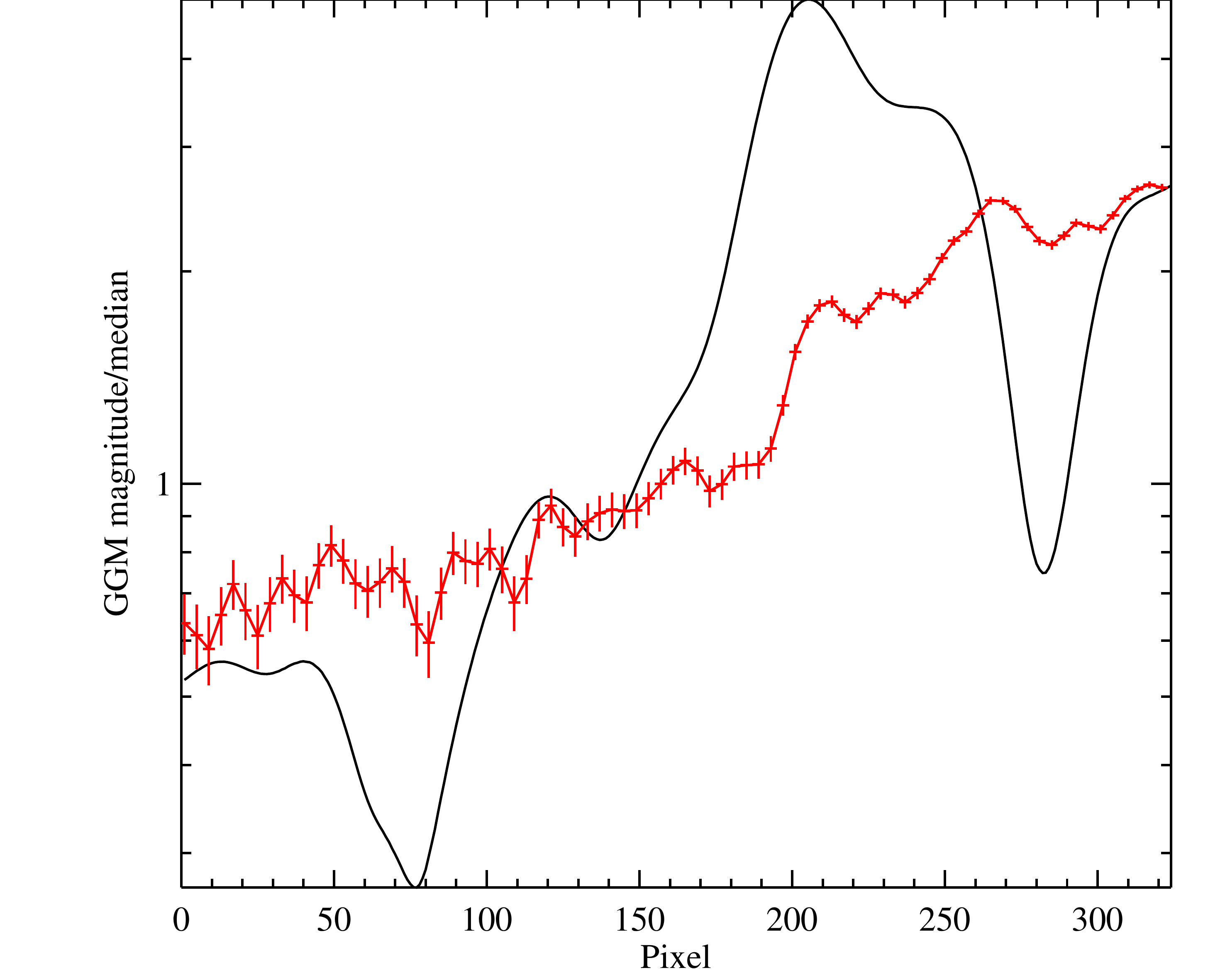}
             
   }
    \hbox{     
          \includegraphics[width=0.6\linewidth]{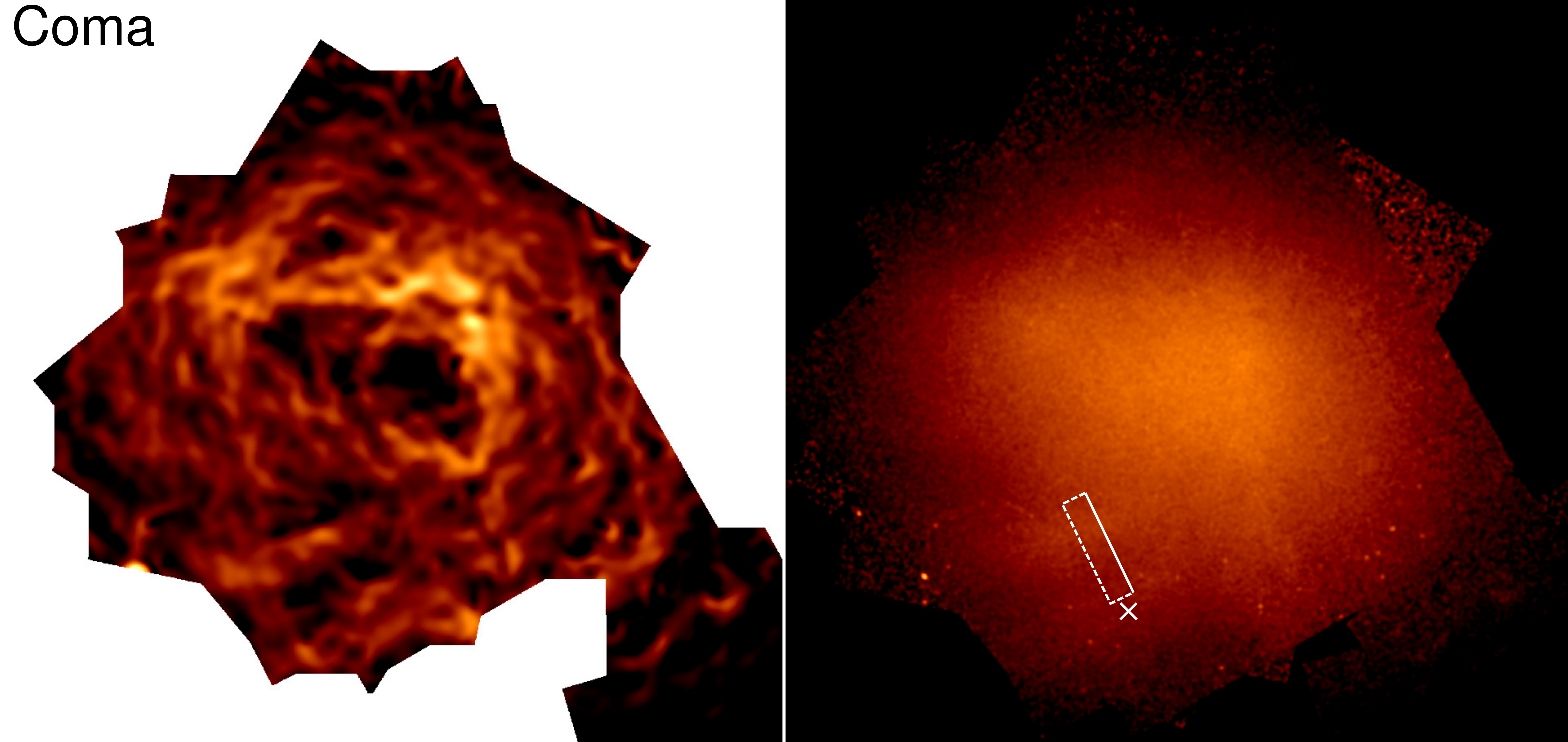}
     \includegraphics[width=0.3\linewidth]{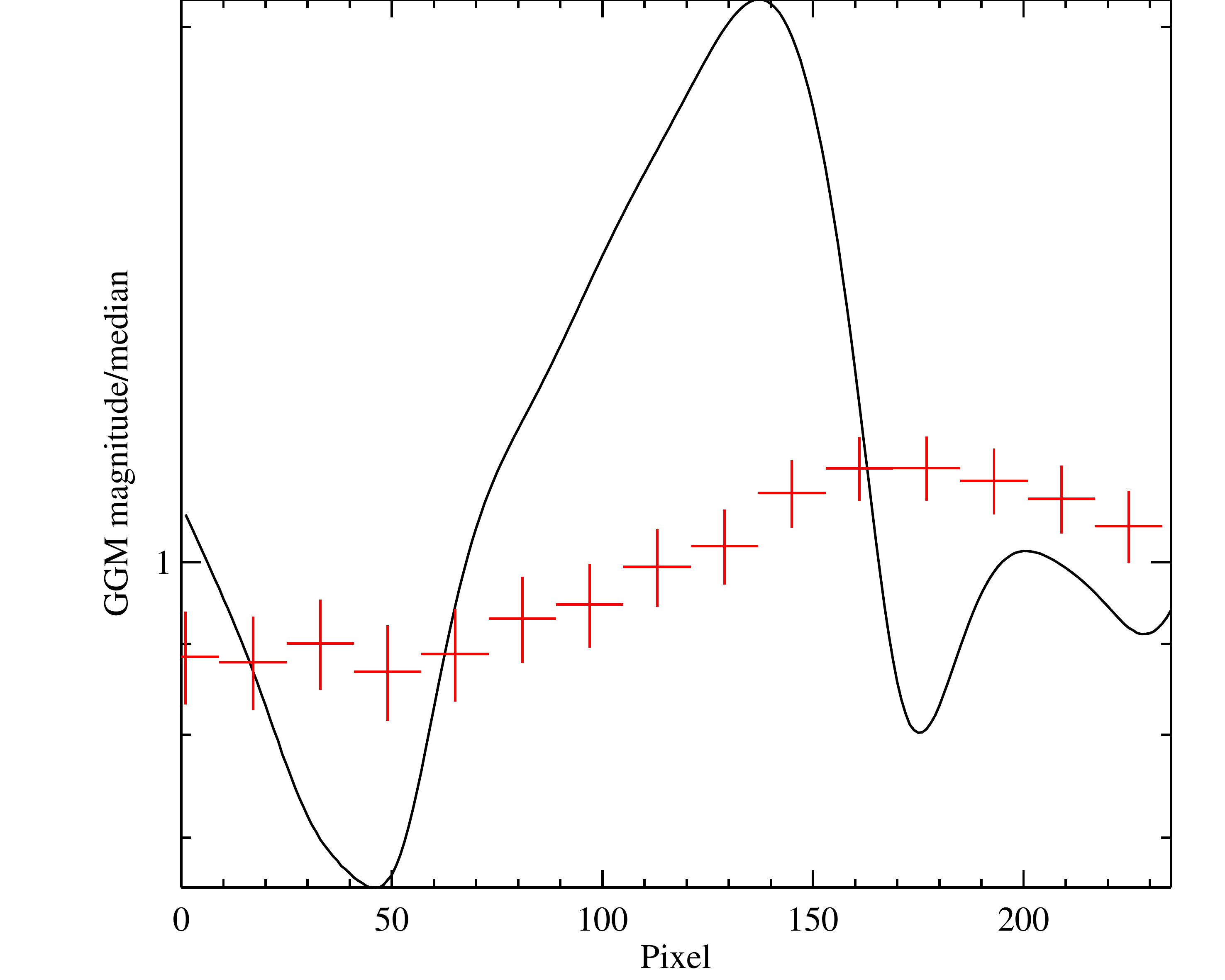}

   }

      \caption{Continuation of Fig. \ref{appendixfigure}}
      \label{appfig2}
  \end{center}
\end{figure*}

\begin{table*}
\begin{center}
\caption{Chandra and XMM-Newton observations used}
\leavevmode
\begin{tabular}{ ll } \hline \hline
Object & Obsids  \\ \hline
Abell 2142 & 15186,16564,16564  \\
Abell 496 & 4976   \\
Crab & 13139,13150,13151,13152,13153,13154  \\
     & 13204,13205,13206,13207,13208,13209 \\
     & 13210,13750,13751,13752,14416,13754 \\
     & 13755,13756,13757,14458,14678,14679 \\
     & 14680,14681,14682,14685,16245,16257 \\
     & 16258,16357,16358,16359 \\
Bullet cluster & 4984,4985,4986,5355,5356,5357,5358,5361  \\
 NGC 5813 & 12951,12952,12953,13246,13247,13253 \\
&  13255,5907,9517  \\
Hercules A & 5796,6257  \\
PKS1404 & 12884  \\
Abell 2052 & 10477,10478,10479,10480,10879,10914,10915 \\
& 10916,10917,5807  \\
Abell 2319 & 15187  \\ 
Coma & 13993,14410,13994,14411,13995,14406,14415  \\ 
& 13996    \\ 
Abell 3667 & 513,5751,5752,5753,6292,6295,6296 \\
Perseus (XMM) & 0204720101,0204720201,0305690301,0305690401 \\
     & 0305720301,0405410101,0673020201,0673020301 \\
    & 0673020401 \\ \hline
\label{alldata}
\end{tabular}
\end{center}
\end{table*}

\end{document}